\documentclass[11pt,a4paper]{article}
\pdfoutput=1

\usepackage{jcappub}

\usepackage{epsf,epsfig}
\usepackage{graphics}
\usepackage{color}

\newcommand{\beq}{\begin{equation}}
\newcommand{\eeq}{\end{equation}}
\newcommand{\bea}{\begin{eqnarray}}
\newcommand{\eea}{\end{eqnarray}}
\newcommand{\vc}[1]{{\textbf{#1}}}
\newcommand{\bmatr}{\begin{pmatrix}}
\newcommand{\ematr}{\end{pmatrix}}

\newcommand{\la}{\langle}
\newcommand{\ra}{\rangle}

\newcommand{\lsim}
{\;\raisebox{-.3em}{$\stackrel{\displaystyle <}{\sim}$}\;}

\newcommand{\red}{\color{red}}

\title{Functional renormalization group for stochastic inflation}

\author{Tomislav Prokopec$^a$ and}
\affiliation{$^a$ Institute for Theoretical Physics, Spinoza Institute and the
	Center for Extreme Matter and Emergent Phenomena (EMMEΦ),
	Utrecht University, Buys Ballot Building,
	Princetonplein 5, 3584 CC Utrecht, the Netherlands}
\author{Gerasimos Rigopoulos$^b$}
\affiliation{$^b$   School of Mathematics, Statistics and Physics, 
	Herschel Building, Newcastle University, 
	Newcastle upon Tyne, NE1 7RU, UK}

\abstract{We apply the functional renormalization group to Starobinsky's stochastic equation describing the local dynamics of a light scalar field in de Sitter. After elaborating on the over-damped regime of stochastic dynamics, we introduce an effective average action for the stochastic field, resulting by progressively integrating out frequencies, and study its flow equation in the local potential approximation (LPA). This effective action determines the approach to equilibrium and allows for the computation of unequal time correlators $\left\langle\phi(t)\phi(t+\Delta t)\right\rangle$ for large values of $\Delta t$. The stochastic RG flow in the LPA can be formulated in two ways, one that preserves the stochastic supersymmetry and one that breaks it. We show that both predict a characteristic decay time very close to that determined by the dynamical mass for a massless self-interacting scalar in de Sitter $m^2\sim \sqrt{\lambda}H^2$. Furthermore, the \emph{temporal} supersymmetric formulation remarkably recovers the flow for the effective potential found using Quantum Field Theory methods and a smoothing over \emph{spatial} wavelengths. We also discuss how the stochastic framework generically predicts an infrared mass which is a few percent smaller than the dynamical mass obtained in the LPA. Our results further support the notion that stochastic inflation captures the correct IR dynamics of light scalar fields in inflation.}

\begin{document}
\maketitle

\section{Introduction}

It is by now well accepted that long wavelength vacuum fluctuations of light scalar fields in inflation can be described via a local Langevin equation, where the field evolves under the influence of a classical stochastic force and a corresponding Fokker-Planck equation. This insight was first developed in the pioneering work of Starobinsky more than 30 years ago \cite{Starobinsky:1986fx} and since then a very large number of works, too numerous to mention here, have utilised it to develop computational tools for inflationary predictions. 

It has also been known for a long time that a massless scalar theory 
does not possess a de Sitter invariant propagator~\cite{Allen:1987tz}. When viewed as a real time evolution in the stochastic framework, this is signified by the unrestrained growth of the field variance with time $\langle\phi^2\rangle$ which is because the IR field at a given point in space performs an unrestrained random walk. This poses problems for pertubative evaluations in the more physically relevant case of an interacting field which inherits an ill-defined perturbation expansion that diverges at late times - see \cite{Seery:2010kh} and references therein for a review of IR effects and associated divergences in inflation. However, even a \emph{non-zero} mass that is not sufficiently large causes trouble as the perturbation series does not apparently converge. Indeed, perturbation theory gives to two loops\footnote{In this paper we write the quartic interation as $\frac{\lambda}{4}\phi^4$. } 
\beq\label{series}
\lim\limits_{t\rightarrow\infty} \langle\phi(t)^2\rangle = \frac{3H^4}{8\pi^2m^2}\left(1-\frac{3}{2}\frac{3\lambda H^4}{4\pi^2 m^4} + 6\left(\frac{3\lambda H^4}{4\pi^2 m^4}\right)^2 + \ldots \right)
\,,
\eeq 
where $H$ denotes the Hubble parameter and $m$ the scalar field mass.
That means that, in a massive scalar theory on de Sitter, the relevant perturvative parameter is not $\lambda$ but instead $\lambda H^4/m^4$ \cite{Burgess:2010dd, Garbrecht:2013coa}, and perturbation theory applies only when $\lambda H^4/m^4\ll 1$. The enhancement 
factor $H^4/m^4$ 
comes from integrating out all the particles that circle in the quantum loops on de Sitter.
We see that even if the field's mass is not zero and the tree-level two-point function is well defined in de Sitter, it is not obvious that higher order loop corrections in perturbation theory lead to a finite result for arbitrarily small masses or, of course, for relatively strong coupling.

Stochastic dynamics allows the re-summation of the series (\ref{series}) to a well defined result for the field variance at late times $\lim\limits_{t\rightarrow \infty}\langle\phi(t)^2\rangle$, as first pointed out by \cite{Starobinsky:1994bd}. In fact, the IR fluctuations can be seen to induce a ``dynamical mass'' $m_{\rm dyn}^2 \sim \sqrt{\lambda}H^2$ for the field \cite{Burgess:2009bs, Garbrecht:2011gu, Serreau:2011fu, Serreau:2013eoa, Guilleux:2015pma, Guilleux:2016oqv}. Euclidean de~Sitter space computations \cite{Rajaraman:2010xd, Beneke:2012kn, Nacir:2016fzi} also support this although they purport to equilibrium and do not provide a dynamical context.  

Here we examine another aspect of mass generation in de Sitter. We are interested in the \emph{temporal correlator} $\lim\limits_{\Delta t\rightarrow \infty}\langle\phi(t+\Delta t)\phi(t)\rangle$ which also perturbatively diverges as there is no mass to suppress long time fluctuations. 
Note that this is an observable quantity as one can in principle perform local measurements on de Sitter that correspond to the correlator
$\langle\phi(t+\Delta t)\phi(t)\rangle$ for large time separations, {\it i.e. } for $\Delta t\gg 1/H$, where 
$1/H$ is the Hubble time. These fluctuations are accessible to a sufficiently long lived local observer during inflation as they involve field measurements for points separated by physical spatial distances less than the Hubble radius.  
This is in notable contrast with earlier work on the functional renormalization group in de Sitter \cite{Serreau:2013eoa, Guilleux:2015pma, Guilleux:2016oqv, Kaya:2013bga} which smooths in space and the effective action flows with increasing wavelength. Such correlations are indirectly accessible to observers only in a post inflationary epoch. As we will see below stochastic inflation and temporal smoothing can recover a flow equation identical to that obtained via spatial coarse graining, see (\ref{RG flow}).  This should probably not come as a surprise, since 
it follows from de Sitter invariance of the correlators (for a discussion of the tree level correlations      
see Eqs.~(11) and~(27) of~\cite{Garbrecht:2013coa}).

A heuristic argument using stochastic inflation \cite{Riotto:2008mv, Garbrecht:2014dca} shows that the generated mass has dynamical consequences and defines a timescale over which long time correlations decay.
Using Eq.~(\ref{eq:starobinsky}) evaluated at the time $t$, multiplying with $\phi(0)$ and taking
the expectation value we have
\begin{align}
\frac{d}{dt}\langle\phi(t)\phi(0)\rangle+\frac{\lambda}{3H}\langle\phi^3(t) \phi(0)\rangle\approx 0\,,
\end{align}
where we assume that $t$ is large enough for the field and the noise to be uncorrelated,
$\langle\xi(t)\phi(0)\rangle\approx 0$. Approximating the correlation of four fields through a Wick expansion we obtain
\begin{align}
\label{unequaltime:gauss}
\frac{d}{dt}\langle \phi(t)\phi(0)\rangle\approx -\frac{\lambda}{H}\langle \phi^2(t)\rangle\langle\phi(t)\phi(0)\rangle\,.
\end{align}
A known result from stochastic inflation \cite{Starobinsky:1994bd} would then allow us to compute 
\beq \label{static}
\lim\limits_{t\rightarrow \infty} \langle\phi^2(t)\rangle = \sqrt{\frac{3}{2}} \frac{\Gamma\left(\frac{3}{4}\right)}{\Gamma\left(\frac{1}{4}\right)}\frac{H^2}{\pi\sqrt{\lambda}}\simeq 0.132\frac{H^2}{\sqrt{\lambda}}\,,
\eeq 
which leads to the estimate
\begin{align}
\langle \phi(t)\phi(0)\rangle\approx \langle\phi^2(0)\rangle{\rm e}^{- 0.132 \sqrt \lambda H t}\,.
\label{rough estimate}
\end{align}
Comparing with the standard result for a massive, light field in de Sitter
\beq
\langle\phi^2 \rangle = \frac{3H^4}{8\pi^2m^2} \,,
\eeq
the quantity 
\beq\label{mdyn}
m^2_{\rm dyn} = \frac{\sqrt{6}}{8\pi}\frac{\Gamma(1/4)}{\Gamma(3/4)}\sqrt{\lambda_{k=H}}H^2\simeq 0.2884\sqrt{\lambda_{k=H}}H^2\,
\eeq
is often referred to as a generated ``dynamical mass'', where $\lambda_{k=H}$ is the ``bare'' coupling on timescales $H^{-1}$, uncorrected from any long-time fluctuations.  

In this work we address these long time correlations 
in de~Sitter by applying the functional renormalization group \cite{Berges:2000ew, Delamotte:2007pf} to the stochastic dynamics. We perform our analysis in the local potential approximation (LPA) considering two cases of regulator function, one that explicitly preserves the underlying supersymmetry of the stochastic action and one that breaks it. The supersymmetry preserving case results in a flow equation which exactly predicts $m^2_{\rm IR}=m^2_{\rm dyn}$ defined in (\ref{mdyn}) \cite{Guilleux:2015pma}. We find that the supersymmetry breaking approach results in an effective mass
\beq\label{mIR}
m^2_{\rm IR} \simeq m^2_{\rm dyn} \approx  0.2784 \sqrt{\lambda_{k=H}}H^2  \,.
\eeq
which is in pretty close agreement with the above result. 

As was pointed out in \cite{Starobinsky:1994bd} the stochastic framework allows for the decay of temporal correlators to be computed without resource to an RG analysis. We find that the decay of temporal correlators is in fact determined by 
\beq
m^2_{\rm stoch}= 0.2668 \sqrt{\lambda_{k=H}}H^2\,,
\eeq     
confirming the earlier result of \cite{Starobinsky:1994bd}. Therefore, correlations decay somewhat more slowly in time than suggested by the value of $m^2_{\rm dyn}$ by a factor of about 8\%. Such a difference between $m^2_{\rm IR}$ and $m^2_{\rm dyn}$ was first discussed in \cite{Gautier:2013aoa, Gautier:2015pca} using QFT computations in the large N limit. We see that stochastic inflation qualitatively reproduces this conclusion for $N=1$ directly, although precise numerical comparison would be beyond the scope of this paper. To recover this result in the RG framework one probably needs to go beyond the LPA.\footnote{We thank Julien Serreau for pointing this out.}

In this paper we also examine the symmetry breaking case $m^2<0$ and find that if the system is followed over long timescales ($k\rightarrow 0$), fluctuations do restore the symmetry when interactions are strong enough; the RG flow generates a positive mass $m^2_{\rm IR}$ also given by (\ref{mIR}). Again, this mirrors the QFT results of \cite{Serreau:2011fu, Lazzari:2013boa, Serreau:2013eoa,Gonzalez:2016jrn}. In our case the following interpretation for the symmetry restoration presents itself: Strong fluctuations can take the field above the barrier into the opposite minimum and back. If the behaviour of the field is averaged over timescales larger than the characteristic time of such jumps, the field does not exhibit preference to be in either of the symmetry breaking states and effectively the symmetry is restored. 

We believe these correspondences further support the proposition that stochastic inflation captures completely and correctly the IR dynamics of quantum fields in de Sitter (for a proof to all orders in perturbation theory see~\cite{Tsamis:2005hd}). In passing we note that it is now known how to stochastisize Yukawa~\cite{Miao:2006pn} and scalar electrodynamics on de Sitter~\cite{Prokopec:2007ak,Prokopec:2008gw,Prokopec:2006ue}, but quantum gravity
with matter is still beyond reach~\cite{Tsamis:2005hd} (for recent attempts in this direction see~\cite{Vennin:2015hra,Pattison:2017mbe,Miao:2018bol}).

The Functional Renormalization Group equation (\ref{RG flow}) is derived by explicitly preserving the supersymmetry of our stochastic action \cite{Parisi:1979ka} and requires the supersymmetry-preserving adaptation of the functional RG technology developed in \cite{Synatschke:2008pv}. This flow equation is mathematically identical to that obtained in \cite{Guilleux:2015pma}, see also \cite{Kaya:2013bga}. It is important to note however that it has been arrived at in a completely different context. The authors of \cite{Guilleux:2015pma} derive it based on field theory and a smoothing of spatial scales. Our derivation is based on spatially local stochastic theory and a smoothing of temporal scales. For us $k$ is frequency while for Serreau et al $\kappa$ is wavenumber. It is rather remarkable that we get an identical flow equation from such a different perspective. Albeit we have no complete understanding of the reasons behind this fact, it probably boils down to de Sitter symmetry being preserved by the stochastic dynamics at late times. We further offer a simple approximation to the solution of the supersymmetric RG flow equation in order to explore mass generation and symmetry restoration in a simple fashion semi-analytically. Our approximation is crude but allows for a simple understanding of the features described above. We leave a more refined analysis of the flow equation, such as that performed in \cite{Lazzari:2013boa}, for future work.

\section{Stochastic inflationary dynamics}

We will consider the IR dynamics of a quantum field $\phi$ in de~Sitter with second-order-in-time equations of motion. As discussed in \cite{Moss:2016uix, Rigopoulos:2016oko}, a separation of long and short wavelengths in the path integral via the propagator leads to a stochastic equation in phase-space, also known as a Kramers equation: 
\beq\label{Kramers}
\partial_t{W}=
\left[\frac{9H^5}{8\pi^2}\frac{\partial^2}{\partial v^2}+
3H\frac{\partial}{\partial v}v+\frac{dV_{\rm eff}}{d \phi}
\frac{\partial}{\partial v}-
v
\frac{\partial}{\partial\phi}
\right]{W}\,
\eeq
instead of the more commonly used Fokker-Planck type
\beq\label{FP1}
\partial_t{P}= \frac{\partial}{\partial\phi}\left(\frac{H^3}{8\pi^2}\frac{\partial}{\partial\phi}+
\frac{dV_{\rm eff}}{d\phi}\frac{1}{3H}\right)P\,.
\eeq
Here $W(v,\phi, t)$ is the probability density for both $\phi$ and its velocity $v$ as a function of time while the probability density $P(\phi,t)$ only depends on $\phi$ and $t$.  $V_{\rm eff}(\phi)$ is the effective
potential calculated at integrating the field fluctuations above the Hubble scale. For notational simplicity, from now on 
we shall denote $V_{\rm eff}(\phi)$ simply by $V(\phi)$.

In this section we recall that (\ref{Kramers}) corresponds to a Langevin equation with the noise acting on the velocity and not the field variable.\footnote{This is therefore different from the quantum phase space approach of \cite{Habib:1992ci}.} We further show how to reduce (\ref{Kramers}) to an equation for $P$ which reproduces the familiar Fokker-Planck equation at leading order in $V''/H^2$. An operator based reduction of this kind can be found in \cite{Risken}; we here re-derive those results using path integral methods. The reduction of the full phase space dynamics to that involving only $\phi$ is possible since $v$ evolves on a timescale of $H^{-1}$ while the slower dynamics takes place on timescales set by $H/m^2$.  

\subsection{Effective IR theory}

We consider a light scalar field minimally coupled to gravity. If all modes below a fixed physical scale $\Delta r \sim 1/H$ are integrated out, the resulting theory has the following effective action per horizon volume \cite{Rigopoulos:2016oko, Moss:2016uix}\footnote{Briefly, this is achieved by splitting the propagator into a IR smoothed part and a short wavelength part and integrating out fluctuations governed by the short wavelength correlator.}
\beq\label{action1}
 S_{\rm eff}[\phi,\psi] = -\int dt \, \psi\left(\ddot{\phi} +3H \dot{\phi} + \frac{d V}{d\phi}\right) +i\, \frac{1}{2}\frac{9H^5}{4\pi^2} \int dt \,\psi^2\,,  
\eeq      
describing temporal dynamics of the field in a single Hubble patch. The potential $V$ is the effective potential resulting from integrating out subhorizon fluctuations. Any correlator of long wavelength fields at fixed physical separation $\Delta r < H^{-1}$ is to be computed with $e^{iS_{\rm eff}}$ as weight
\beq
\big\langle\ldots \big \rangle = \int D\phi D\psi \,\ldots\, e^{iS_{\rm eff}}\,.
\eeq
The field $\psi$ is related to the difference field $\phi_+-\phi_-$ of the Schwinger-Keldysh formalism 
and signifies the amplitude of quantum fluctuations around the mean field value,
while $\phi$ relates to the average field, $\phi=(\phi_+ +\phi_-)/2$, and signifies 
expectation value of the field operator $\langle\hat \phi\rangle$. 

By introducing an auxiliary field $z$ we can write the action in terms of first order derivatives 
\beq
S[\phi,z,v,\psi] = - \int dt \left[z\left(\dot{\phi}-v\right)+\psi\left(\dot{v} + 3Hv+\frac{dV}{d\phi}\right) \right]  +i\, \frac{1}{2}\frac{9H^5}{4\pi^2} \int dt \,\psi^2\,.
\eeq  
To see the type of dynamics the above action describes it is convenient to perform a Hubbard-Stratonovich transformation and write
\beq
e^{-\frac{1}{2} \int dt\, \frac{9H^5}{4\pi^2} \psi^2} = \int D\xi \, e^{-\int dt\,\frac{1}{2}\frac{4\pi^2}{9H^5}\xi^2 + i\int dt\,\xi\psi } 
\,,
\eeq
which implies that the action can be written as 
\beq
S[\phi,z,v,\psi] = - \int dt \left[z\left(\dot{\phi}-v\right)+\psi\left(\dot{v} + 3Hv+\frac{dV}{d\phi} - \xi\right) \right]  
\eeq  
and $\xi$ is a Gaussian random field with 
\beq\label{noise1}
\langle\xi(t)\xi(t')\rangle = \frac{9H^5}{4\pi^2}\delta(t-t')\,.
\eeq
The above action corresponds to Langevin dynamics 
\bea\label{2nd order stoch}
\dot{v}+3Hv + \frac{dV}{d\phi}=\xi\,,\\
\dot{\phi}=v\,,
\eea 
provided that the time derivative is interpreted in a retarded, or Ito, sense - see e.~g.~\cite{ZinnJustin:2002ru, Kleinert, Kamenev} and the appendix. We see that in the above approach the stochastic noise emerges as a proper \emph{force} imparting acceleration rather than a stochastic displacement of the field value as the process is usually described. This is ultimately related to the fact that the field $\psi$ responsible for the stochastic term is conjugate to $v$ and not $\phi$. In the next subsection we derive the more common over-damped dynamics, valid when evolution on timescales $\Delta t \sim 1/H$ are ignored. The over-damped limit is also known as the slow roll regime which implies $V''/H^2\ll 1$ \cite{Risken} and the derivation below can also be thought of as an expansion in this parameter.            

It is also worth noting that the second order stochastic dynamics implied by (\ref{action1}) 
\beq\label{stoch-second}
\ddot{\phi}+3H\dot{\phi}+\frac{dV}{d\phi}=\xi\,,
\eeq 
along with (\ref{noise1}), can be obtained from the more commonly employed split  
\bea
\dot{v}+3Hv + \frac{dV}{d\phi}=\xi_v\,, \label{v-stoch}\\
\dot{\phi}=v+\xi_\phi \label{phi-stoch}\,,
\eea    
with an appropriate choice of window function $w$.
The noise terms are schematically defined as 
\beq
\xi_v(t, \vc{x}) = \int d^3k \, { \dot{w}}(k,t)\,\dot{q}_{\vc{k}}(t)\,e^{i\vc{k}\vc{x}}\,,\quad \xi_\phi(t,\vc{x})= \int d^3k  \, { \dot{w}}(k,t)\,q_{\vc{k}}(t)\,e^{i\vc{k}\vc{x}}
\eeq 
where $q_\vc{k}$ includes the linear mode function and 
Gaussian random fields that stand for the creation and annihilation operators. 
Taking a derivative of (\ref{phi-stoch}) and substituting (\ref{v-stoch}) one obtains
\beq
\ddot{\phi} + 3H \dot{\phi} + \frac{d V}{d \phi} = \int d^3k \left[\left( \ddot{w} + 3H\dot{w}\right)q_{\vc{k}}+2\dot{w}\dot{q}_{\vc{k}}\right] e^{i\vc{k}\vc{x}} \,.
\eeq    
Ignoring the subdominant $\dot{q}_{\vc{k}}$ term and choosing the window function to be \cite{Rigopoulos:2016oko}
\beq\label{window}
w_k(t)=\left(1-\frac{k^3}{(\epsilon a H)^3}\right)\,\Theta\!\left[\ln\left(\frac{\epsilon a H}{k}\right)\right]%=1-\left({1-e^{-3H\left(t-\frac{1}{H}\ln\left(k/\epsilon H\right)\right)}}\right)\,\,\Theta\left(t-\frac{1}{H}\ln\left(k/\epsilon H\right)\right)
\eeq
we see that $\ddot{w}_k(t)+3H\dot{w}_k(t)=3H\delta(t-\frac{1}{H}\ln\left(k/\epsilon H\right))$ and one obtains the second order stochastic equation (\ref{stoch-second}).

\subsection{Over-damped limit}
Before proceeding with our renormalization group analysis we discuss how the Kramers equation (\ref{Kramers}) relates to the more standard Fokker-Planck equation in the over-damped (slow roll) limit. We do this in two ways in the following sections, first by working directly with the Kramers equation and secondly by utilizing the path integral formulation.      

\subsubsection{Fokker-Planck limit}
Let us first change phase space coordinates by subtracting the slow roll ``drift'' velocity
\beq
\tilde{v}=v+\frac{1}{3H}\frac{dV}{d\phi}
\eeq 
and the Kramers equation for the transformed probability density  $\tilde{W}=\tilde{W}(t,\tilde{v},\phi)$
becomes
\beq\label{Kramers 2}
\left[\frac{\partial}{\partial t}+\tilde{v}\frac{\partial}{\partial \phi}-\frac{1}{3H}\frac{\partial}{\partial\phi}V'-\frac{V'V''}{9H^2}\frac{\partial}{\partial\tilde{v}}\right]\tilde{W} = 3H\frac{\partial}{\partial \tilde{v}}\left[\left(1-\frac{V''}{9H^2}\right)\tilde{v} + \frac{3H^4}{8\pi^2}\frac{\partial}{\partial\tilde{v}}\right]\tilde{W}\,.
\eeq
Defining the integrated probability density and current
\beq
P(\phi,t)=\int d\tilde{v} \, \tilde{W}(\tilde{v},\phi,t)  \,,\quad J(\phi,t)= \int d\tilde{v}\, \tilde{v}\, \tilde{W}(\tilde{v},\phi,t)
\eeq
one obtains
\beq\label{density}
\frac{\partial P}{\partial t} =-\frac{\partial}{\partial\phi} \left(J-P\frac{V'}{3H}\right)
\eeq
and 
\beq\label{current}
\frac{dJ}{dt} \equiv\left(\frac{\partial }{\partial t} - \frac{V'}{3H}\frac{\partial }{\partial \phi}\right) J= -2\frac{\partial K}{\partial\phi} + \frac{V''}{3H}J - 3H\left(1-\frac{V''}{9H^2}\right) J - \frac{V'V''}{9H^2}P
\eeq
where the lhs of (\ref{current}) denotes the convective derivative along the slow-roll flow and 
\beq
K=\int d\tilde{v}\,\frac{\tilde{v}^2}{2}\,\tilde{W}(\tilde{v},\phi,t)\,.
\eeq
Dropping subdominant, slow-roll suppressed terms and assuming that in the overdamped limit $3HJ$ is large compared to the convective derivative on the lhs, we have that the {\it Ansatz} 
\beq\label{phi-v distn}
\tilde{W}\simeq P(\phi,t)\frac{\exp\left[-\frac{1}{2}\frac{\tilde{v}^2}{3H^4/8\pi^2}\right]}{\sqrt{2\pi}\sqrt{ 3H^4/8\pi^2}}
\eeq
leads to
\beq
3H J\simeq-\frac{3H^4}{8\pi^2}\frac{\partial P}{\partial\phi}\,.
\eeq
Plugging this in (\ref{density}) we obtain the Fokker-Planck  equation for $P$
\beq\label{FP2}
\partial_t P = \left(\frac{1}{2}\frac{H^3}{4\pi^2}\partial_\phi^2 + \partial_\phi\, \frac{V'}{3H}\right)P\,.
\eeq
If the system starts away from the equilibrium position at the minimum of the potential it evolves towards it with a small spread in velocities around the classical slow roll solution $v=-V'/3H$ and a $\phi$ distribution determined by (\ref{FP2}). When the field reaches the minimum of the potential where $V'=0$, we obtain the result of \cite{Rigopoulos:2016oko, Moss:2016uix} for the velocity distribution. Hence, the velocity of the field cannot be determined to better accuracy than that implied by (\ref{phi-v distn}).

\subsubsection{Path Integral} To obtain an approximation to the second order stochastic dynamics on timescales above $\Delta t \sim H^{-1}$, we can start by explicitly performing the $\psi$ integral   
\beq
\int D\psi \,\,e^{-\int dt \, \left[\frac{1}{2} \frac{9H^5}{4\pi^2}\psi^2  +i\psi\left( \dot{v}+3Hv+V'\right)\right]} = e^{-\int dt\,\frac{1}{2}\frac{4\pi^2}{9H^5}\left( \dot{v}+3Hv+V'\right)^2}
\eeq
which brings the exponent in the form (up to boundary terms)
\bea
&&iS[\phi,z,v]=\nonumber\\ &=& \int dt \left\{-i z\dot{\phi}-\frac{1}{2}\frac{4\pi^2}{9H^5}V'^2- \frac{1}{2}\frac{4\pi^2}{9H^5} \left[v\left(-\partial_t^2+9H^2\right)v +v\left(-i\frac{9H^5}{2\pi^2}z+6HV'-2V''\dot{\phi}\right) \right] \right\}\,. \nonumber%\\
%&&-\frac{1}{2}\frac{4\pi^2}{9H^5}\left[\frac{3H}{2}v^2+v\dot{v}+2vV'\right]^t_0
\eea
We can now perform the Gaussian $v$ path integral (again up to boundary terms). 
\bea
\int D v \,  e^{iS} &=& \exp\bigg[
%-\frac{1}{2}\frac{4\pi^2}{9H^5}\left[\frac{3H}{2}v^2+v\dot{v}+2vV'\right]^t_0-\mathcal{A}_{\rm cl}-\frac{1}{2}{\rm Tr}{\ln} \left(-\partial_t^2+9H^2\right) \nonumber\\
%&&
+ \int dt \left\{-i z\dot{\phi}-\frac{1}{2}\frac{4\pi^2}{9H^5}V'^2\right\}+\frac{1}{2}\frac{\pi^2}{9H^5}\int dtdt'\mathcal{A}(t)G(t,t')\mathcal{A}(t')\bigg]\,,
\eea
where 
\beq
\mathcal{A}(t)=-i\frac{9H^5}{2\pi^2}z+6HV'-2V''\dot{\phi}
\eeq
and
\beq
\left(-\partial_t^2+9H^2\right)G(t,t') =\delta(t-t')\,.
\eeq
For our purposes we can approximate $G(t,t')\simeq \frac{1}{9H^2}\delta(t-t')$ and the exponent becomes
\bea\label{action2}
iS[z,\phi] &\simeq& \int dt \left(-i z\dot{\phi}-\frac{1}{2}\frac{4\pi^2}{9H^5}V'^2\right)
+\frac{1}{2}\frac{\pi^2}{81H^7}\int dt\left(-i\frac{9H^5}{2\pi^2}z+6HV'-2V''\dot{\phi}\right)^2\nonumber\\
&\simeq&  
\int dt \left\{-\frac{1}{2}\frac{H^3}{4\pi^2}z^2-iz\left[\left(1-\frac{V''}{9H^2}\right)\dot{\phi}+\frac{V'}{3H}\right]\right\} \, ,
\eea  
where in the last approximate equality we ignored a term of $\mathcal{O}\left[\left(\frac{V''}{ 9H^2}\right)^2\right]$. After setting $z=u\left(1+\frac{V''}{9H^2}\right)$ we can write
\beq
iS(u,\phi)\simeq\int dt\left\{-\frac{1}{2}\frac{H^3}{4\pi^2}\left(1+2\frac{V''}{9H^2}\right)u^2+iu\left[\dot{\phi}+\left(1+\frac{V''}{9H^2}\right)\frac{V'}{3H}\right]\right\}\,.
\eeq 
We therefore find that the over-damped limit corresponds to the Langevin equation
\beq
\dot{\phi} + \left(1+\frac{V''}{9H^2}\right)\frac{\partial_\phi V}{3H} = \xi(t),
\label{eq:starobinsky+}
\eeq
with the stochastic noise governed by a gaussian probability distribution functional for a noise history $\xi(t)$
\beq
P[\xi]\propto \exp\left(- \frac{1}{2} \int dt\, \frac{4 \pi^2}{H^3\left(1+2\frac{V''}{9H^2}\right)}\xi(t)^2 \right)
\eeq             	
implying white noise but with a field dependent amplitude  
\beq
\la \xi(t)\xi(t^\prime) \ra = \frac{H^3}{4\pi^2}\left(1+2\frac{V''}{9H^2}\right)\delta(t-t^\prime)\,.
\label{eq:noise_correlation+}
\eeq

We have shown how the stochastic dynamical (\ref{2nd order stoch}) reduces to the standard Starobinsky-Langevin equation, modified by corrections of order $\frac{V''}{H^2}$. These corrections are also derived in \cite{Risken} via a more conventional but laborious operator-based approach. Therefore, memory of the second order nature of the system is retained, even in the over-damped  limit, in these sub-dominant corrections. This observation may be interesting for inflationary perturbations and we hope to return to it in a future publication. However, they somewhat complicate the RG analysis that follows. To keep things simpler, we proceed with    
\beq\label{stoch path}
iS\simeq -\int dt\left[ \frac{1}{2}\frac{H^3}{4\pi^2}u^2-iu\left(\dot{\phi}+\frac{V'}{3H}\right)\right]\,,
\eeq 
which can be associated with the stochastic Starobinsky dynamics, 
\beq
\dot{\phi} + \frac{\partial_\phi V}{3H} = \xi(t),
\label{eq:starobinsky}
\eeq
\beq
\la \xi(t)\xi(t^\prime) \ra = \frac{H^3}{4\pi^2}\delta(t-t^\prime)\,.
\label{eq:noise_correlation}
\eeq
The path integral used for obtaining correlation functions w.r.t. stochastic histories driven by $\xi$ is then given by
\beq
\big\langle\ldots \big \rangle = \int D\phi Du\,\ldots\, e^{-\int dt\left[ \frac{1}{2}\frac{H^3}{4\pi^2}u^2-iu\left(\dot{\phi}+\frac{V'}{3H}\right)\right]}P(\phi_{\rm i})
\eeq      
where we have included the initial probability distribution $P(\phi_{\rm i})$ and the stochastic dynamics can bne associated with a  non-hermitian Hamiltonian~\cite{ZinnJustin:2002ru, Kamenev}       
\beq\label{Hamiltonian 1}
\hat{{\mathbf H}}= \frac{1}{2}\,\frac{H^3}{4\pi^2} \,\hat{u}^2 -i\hat{u}\frac{V'}{3H}
\eeq
or, in the field representation
\beq\label{Hamiltonian}
\langle\phi|\hat{{\mathbf H}}|\phi'\rangle = \left(-\frac{1}{2}\frac{H^3}{4\pi^2}\partial_\phi^2-\partial_\phi\, \frac{V'}{3H}\right)\delta\left(\phi-\phi'\right)\,.
\eeq
Therefore, the ``Schr\"{o}dinger" equation corresponding to the path integral we derived in the IR limit is
\beq\label{Fokker Planck}
\partial_t P = -\hat{{\mathbf H}}P = \left(\frac{1}{2}\frac{H^3}{4\pi^2}\partial_\phi^2 + \partial_\phi\, \frac{V'}{3H}\right)P
\eeq
which is again nothing but the well known Fokker-Planck equation (\ref{FP1}) for $\phi$. 

There is a subtlety assosiated with the action (\ref{stoch path}), formally equivalent to the ambiguity of ordering $\hat{u}$ and $V'(\hat{\phi})$ in the Hamiltonian. This can be resolved by recalling that the origin of (\ref{stoch path}) lies in the Schwinger-Keldysh path integral of the full quantum theory. The latter implies that in their time-discretized form, $u$ is computed one step ahead of $\phi$, suggesting the ordering empoyed in (\ref{Hamiltonian}, \ref{Fokker Planck}). However, this also requires Ito calculus when one passes to the continuum limnit. We can correct for that, allowing for example a more convenient midpoint prescription and standard calculus, by including a determinant in the path integral
\beq\label{functional 1}
\big\langle\ldots \big \rangle = \int D\phi Du\,\ldots\, {\rm Det \left[\partial_t + \frac{V''}{3H}\right]} e^{-\int dt\left[ \frac{1}{2}\frac{H^3}{4\pi^2}u^2-iu\left(\dot{\phi}+\frac{V'}{3H}\right)\right]}\,.
\eeq      
This is the form we will be employing form now on, referring the reader to the literature for more details \cite{ZinnJustin:2002ru, Kleinert,  Garbrecht:2014dca, Kamenev, Garbrecht:2013coa}.

%\beq\label{functional 1}
%\left\{- \int dt \left[ \frac{1}{2} \bmatr \phi, & u \ematr \bmatr 0 & i(-\partial_t + \frac{m^2}{3H}) \\  i(\partial_t + \frac{m^2}{3H}) & \frac{H^3}{4 \pi^2} \ematr \bmatr \phi \\ u \ematr + \frac{\partial_\phi V_{\mathrm{int}}}{3H}u\right]\right\} \,,
%\eeq

\subsection{Relation to Euclidean quantum mechanics} 
To simplify our notation we define new rescaled fields $\varphi$, $F$ and a dimensionless potential $U$
\beq
\phi=\frac{H^{3/2}}{2\pi}\varphi \,,\quad u=\frac{H^{3/2}}{2\pi}\left(F+i\dot{\varphi}\right) \,,\quad V= \frac{3H^4}{4\pi^2}U\,.
\eeq
The determinant can be expressed in two ways. Firstly, we can write 
\beq\label{determinant1}
{\rm Det} \left[\partial_t + U_{\varphi\varphi}\right]={\rm Det} \left[\partial_t\right]{\rm Det}\left[1 + \partial_t^{-1} U_{\varphi\varphi}\right] 
\eeq
and ignoring the field independent determinant
\bea 
{\rm Det}\left[1 + \partial_t^{-1} U_{\varphi\varphi}\right] &=& \exp\left\{{\rm Tr}\log \left[1 + \partial_t^{-1} U_{\varphi\varphi} \right]\right\}\nonumber  \nonumber\\ &=&  \exp\left\{{\rm Tr}\log \left[1+\Theta (t-\tau) U_{\varphi\varphi}(\varphi(\tau))  \right]\right\}\,.
\eea 
Upon expanding the logarithm and taking the trace we see that the only term that survives yields   
\beq\label{determinant2}
{\rm Det}\left[1 + \partial_t^{-1} U_{\varphi\varphi}\right]= \exp\left\{+\Theta(0)\int U_{\varphi\varphi} dt  \right\}\,,
\eeq
where $\Theta(x)$ is the Heaviside function. It is convenient to use the Stratonovich prescription and set $\Theta(0)=\frac{1}{2}$
\beq\label{determinant3}
{\rm Det}\left[1 + \partial_t^{-1} U_{\varphi\varphi}\right]= \exp\left\{+\frac{1}{2}\int U_{\varphi\varphi} dt  \right\}\,.
\eeq
The determinant can also be expressed using ghost fields as
\beq
{\rm Det}\left[\partial_t + U_{\varphi\varphi}\right] = \int D\bar{\psi}D\psi  \,e^{\,+\int dt \, \bar{\psi}\left(\partial_t + U_{\varphi\varphi}\right)\psi} \,.
\eeq
We therefore obtain, including the initial probability distribution $P(\phi_{\rm i})$,
\beq
\big\langle\ldots \big \rangle = \int d\varphi_{\rm f} d\varphi_{\rm i}\,e^{-U(\varphi_{\rm f})+U(\varphi_{\rm i})}\int D\phi Du\,\ldots\,e^{-{S}}P(\phi_{\rm i})\,,
\eeq
$\varphi_{\rm f}$ and $\varphi_{\rm i}$ being the values at final and initial times and where 
\beq\label{Action with ghosts}
S=\int dt \, \left[\frac{1}{2}\dot{\varphi}^2- \bar{\psi}\left(\partial_t+U_{\varphi\varphi}\right)\psi +iFU_{\varphi}+\frac{1}{2}F^2\right]\,,
\eeq
or 
\beq\label{Action without ghosts}
S=\int dt \, \left[\frac{1}{2}\dot{\varphi}^2- \frac{1}{2}U_{\varphi\varphi} +iFU_{\varphi}+\frac{1}{2}F^2\right]\,.
\eeq
The two forms of the action are, of course, equivalent. The $F$ field can also be integrated out and one sees that, apart from the ordinary integrations over the boundary fields $\varphi_{\rm i}$ and $\varphi_{\rm f}$, the stochastic process is equivalent to Euclidean quantum mechanics
\beq\label{action-integrated}
\big\langle\ldots \big\rangle= \int d\varphi_{\rm f} d\varphi_{\rm i}\int D\varphi\,\ldots\,e^{-U(\varphi_{\rm f})+U(\varphi_{\rm i})}P(\varphi_{\rm i})\, \exp \left\{-\int dt \left[\frac{1}{2}\dot{\varphi}^2
+{\cal W}\right]\right\}\,.
\eeq  
with a stochastic potential ${\cal W}$ 
\beq\label{W}
{\cal W}=\frac{1}{2}\left[\left({U_{\varphi}}\right)^2-U_{\varphi\varphi}\right]\,. 
\eeq

It is obvious that the coefficients of different $\varphi$ powers in ${\cal W}$
 are not independent but bear certain relationships, inherited from the choice of the ``superpotential'' $U$. Remarkably, this is reflected in a hidden Parisi-Sourlas \emph{supersymmetry} \cite{Parisi:1979ka} which is expressed in the form (\ref{Action with ghosts}) as
\beq 
\delta\varphi=\bar{\epsilon}\psi - \bar{\psi}\epsilon\,,\quad \delta\psi=\left(\dot{\varphi}-iF\right)\epsilon\,\quad \delta\bar{\psi}=\bar{\epsilon}\left(\dot{\varphi}+iF\right)\,,\quad \delta F = i\bar{\epsilon}\dot{\psi} +i\dot{\bar{\psi}}\epsilon\,.
\eeq
where $\epsilon$ and $\bar{\epsilon}$ are arbitrary infinitesimal Grassmann numbers. This symmetry is also related to the fact that the choice of the ambiguous $\Theta(0)$ in (\ref{determinant2}) is immaterial for the computation of correlation functions \cite{Hertz:2016vpy}. For example, choosing an Ito prescription sets the determinant to unity, and correspondingly $\Theta(0)=0$, but replaces a $\dot{\varphi}U_{\varphi} = \dot{U}$ total derivative term in the above treatment by $\dot{\varphi}U_{\varphi}=\dot{U}-\frac{1}{2}U_{\varphi\varphi}$ according to Ito's lemma. The appearance of a $-\frac{1}{2}U_{\varphi\varphi}$ term is also seen in transforming the rescaled Fokker-Planck equation (\ref{Fokker Planck}) (also setting $P\frac{H^{3/2}}{2\pi} \rightarrow P$) 
\beq\label{FP rescaled}
\partial_t P = \frac{1}{2}\partial_{\varphi\varphi}P+\partial_\varphi\left(U_{\varphi}P\right)
\eeq
into a Euclidean Schr\"odinger (diffusive) equation by writing $P(\varphi,t)=e^{-U(\varphi)}\Psi(\varphi,t)$ for which (\ref{FP rescaled}) gives
\bea\label{Schro-Euclidean}
-\frac{\partial}{\partial t}\Psi(\varphi,t) &=&\left(-\frac{1}{2}\frac{\partial^2}{\partial \varphi^2}+ {\cal W(\varphi)}\right) \Psi(\varphi,t)\,,
\eea 
corresponding to a (Hermitian) Hamiltonian  $H_{\rm st} = -\frac{1}{2}p^2+{\cal W}(\varphi)$ and momentum operator $p=-\frac{\partial}{\partial\varphi}$. Its solution can be written
\beq
\Psi(\varphi,t)=\int d\varphi' \, \langle \varphi,t|\varphi',t'\rangle \, \Psi(\varphi',t')\,
\eeq
where the propagator is expressed as the path integral appearing in (\ref{action-integrated}) 
\beq\label{Euclidean Propagator}
\langle \varphi_1,t|\varphi_2,t'\rangle= \langle \varphi_1|e^{-{\red H_{\rm st}}(t-t')}|\varphi_2 \rangle =\int\limits^{\varphi_1}_{\varphi_2} \mathcal{D}\phi \,\, \exp \left\{-\int\limits_{t'}^t dt \left[\frac{1}{2}\dot{\phi}^2+{\cal W}(\phi) \right]\right\}\,.
\eeq  

Equation (\ref{Schro-Euclidean}) possesses a complete, orthonormal set of eigenfunctions $\Psi_n(\varphi)$, allowing for the solution to be written as
\beq\label{eigen-expansion}
\Psi(x,t) = \sum_{n=0}^{\infty} a_n\Psi_n(\varphi)e^{-\Lambda_n t}\,,
\eeq  
where $\Lambda_n$ are the eigenvalues of the Hamiltonian
\bea\label{Schro-eigenvalue}
 \left(-\frac{1}{2}\frac{\partial^2}{\partial \varphi^2}+{\cal W}(\varphi)\right) \Psi_n(\varphi,t) &=& \Lambda_n \Psi_n\,.
\eea 
Noting that (\ref{Schro-eigenvalue}) can be written as 
\beq\label{eqn:pseudo-Schro}
 \frac{1}{2}\left(-\frac{\partial}{\partial \varphi} + U_\varphi \right)\left(\frac{\partial}{\partial \varphi} + U_\varphi \right)\Psi_n(\varphi,t) = \Lambda_n \Psi_n\,,
\eeq
it is clear that the lowest eigenvalue is $\Lambda_0=0$ corresponding to the ``ground state'' 
\beq
\Psi_0(\varphi)= \frac{1}{\sqrt{\mathcal{N}}}e^{-U(\varphi)} \,, 
\eeq
with the normalization constant 
\beq
\mathcal{N}=\int d\varphi \, e^{-2U(\varphi)}\,.
\eeq
Furthermore, we see from (\ref{eigen-expansion}) that the decay rate of unequal time correlators are determined by the energy of the first excited state of the potential ${\cal W}$, \emph{not} the scalar field potential $U$. We will discuss these points further below (see also \cite{Starobinsky:1994bd}).

\subsection{Correlation Functions}
Knowledge of the Euclidean propagator (or kernel) (\ref{Euclidean Propagator}) 
allows for the construction of a variety of field correlation functions. An obvious example that has been discussed widely are correlation functions involving $\varphi_{\rm f}$, the field at the final time. A simple expression can be obtained if initial conditions were set in the far past and equal time correlations are computed in the far future. Taking the double limit of very early and very late times and noting that only the zero eigenvalue of the Euclidean Hamiltonian contributes - see \cite{Ryder} for the corresponding computation in the Lorentzian QFT case - we obtain
\bea
\lim\limits_{\begin{smallmatrix}t\rightarrow\infty \\t'\rightarrow -\infty \end{smallmatrix}}  \langle \varphi_1, t|\varphi_2, t'\rangle &=& \Psi_0(\varphi_1) \langle 0_{\rm out}|0_{\rm in}\rangle\Psi_0(\varphi_2) 
\nonumber \\
%&=&\frac{e^{-U(\varphi_1)}}{\sqrt{\mathcal{N}}}\frac{e^{-U(\varphi_2)}}{\sqrt{\math%cal{N}}}\nonumber\\
&=&\frac{e^{-\left(U(\varphi_1)+U(\varphi_2) \right)}}{{\mathcal{N}}}
\eea 
giving
\beq
\langle O[\varphi_{\rm f}]\rangle = \int d\varphi_{\rm f} \, O[\varphi_{\rm f}]\, \frac{e^{-2U(\varphi_{\rm f})}}{\mathcal{N}}
\eeq
This of course corresponds to the well known fact that at late times an equilibrium distribution is reached   
\beq\label{equilibrium distn}
P_{\rm eq}(\varphi) = \frac{1}{\mathcal{N}}e^{-2U(\varphi)}\,,
\eeq
and provides the weight with respect to which expectation values are computed.

A common unequal time autocorrelator of the field is given by
\beq\label{edge-correlator}
\langle\varphi(t)\varphi(t') \rangle =
\int d\varphi_{\rm t} d\varphi_{\rm t'}\,\varphi_t\varphi_{t'}\, e^{-U(\varphi_{t})+U(\varphi_{t'})}\langle \varphi_{\rm t},t|\varphi_{\rm t'},t'\rangle P(\varphi_{t'})\,.
\eeq  
Obviously, at very large separations the fields become uncorrelated 
\beq
\lim\limits_{\begin{smallmatrix}t\rightarrow\infty \\t'\rightarrow -\infty \end{smallmatrix}} \langle\varphi(t)\varphi(t') \rangle = \int d\varphi_{\rm t}\,\varphi_{\rm t}P_{\rm eq}(\varphi_{\rm t})\, \int d\varphi_{\rm t'} \,\varphi{\rm t'}\,P(\varphi_{t'})\,,
\eeq
and the correlator tends to zero for even probability distribution functions. In the next section we will discuss the behaviour of this correlation function for large but finite time intervals.   

Finally, we can compute unequal time correlators of fields at different times as averages within different stochastic histories, such as
\beq\label{middle-correlator}
\langle T\varphi(t_1)\ldots \varphi(t_n)\rangle = \int d\phi_{\rm f}d\phi_{\rm i} \,e^{-U(\varphi_{\rm f})}e^{+U(\varphi_{\rm i})} \langle \varphi_{\rm f},t| T\varphi(t_1)\ldots \varphi(t_n)|\varphi_{\rm i},t'\rangle  P(\varphi_{\rm i})
\eeq
where $T$ stands for time ordering and 
\beq
\langle \varphi_{\rm f},t| T\varphi(t_1)\ldots \varphi(t_n)|\varphi_{\rm i},t'\rangle =\int\limits^{\varphi_{\rm f}}_{\varphi_{\rm i}} D\varphi \,  \varphi(t_1)\ldots \varphi(t_n)\, e^{-\int_{t_{\rm i}}^{t_{\rm f}} dt \left[\frac{1}{2}\dot{\varphi}^2+W \right]}\,.
\eeq
Unlike correlators such as (\ref{edge-correlator}), the fields in the above expression are evaluated at intermediate times $t'<t_n<\ldots<t_1<t$ and averaged over different noise histories. To compute such expectation values we can consider
\beq\label{stochastic Gen}
\langle \varphi_1,t|\varphi_2,t'\rangle^J=\int\limits^{\varphi_1}_{\varphi_2} \mathcal{D}\phi \,\, \exp \left\{-\int\limits_{t'}^t dt \left[\frac{1}{2}\dot{\varphi}^2+{\cal W(\varphi)} - J\varphi\right]\right\}
\eeq
where a current $J$ coupled linearly to $\varphi$ has been added to the action. Variation wrt $J$ generates correlation functions in the usual way 
\beq
\langle \varphi_1,t|\varphi(t_1)\ldots \varphi(t_n)|\varphi_2,t'\rangle = \frac{\delta^n}{\delta J(t_1)\ldots \delta J(t_n)}\langle \varphi_1,t|\varphi_2,t'\rangle^J\Big|_{J=0}
\eeq
Equilibrium expectation values can be obtained by taking the large time limit, assuming the source only acts between $-T$ and $+T$ and noting again that only the lowest eigenvalue of the Euclidean Hamiltonian contributes \cite{Ryder}. In our case this argument gives
\bea
\lim\limits_{\begin{smallmatrix}t\rightarrow\infty \\t'\rightarrow -\infty \end{smallmatrix}}  \langle \varphi_1, t|\varphi_2, t'\rangle ^J &=& \Psi_0(\varphi_1)\langle 0,+T|0,-T\rangle^J \Psi_0(\varphi_2) \nonumber \\
&=&\frac{e^{-U(\varphi_1)}}{\sqrt{\mathcal{N}}}\frac{e^{-U(\varphi_2)}}{\sqrt{\mathcal{N}}}\langle 0_{\rm out}|0_{\rm in}\rangle^J
\eea
where we hid the T dependence assuming 2T to be large enough to support any source $J(t)$ of interest. Hence, for large times 
\bea
\langle T \varphi(t_1)\ldots \varphi(t_n)\rangle &=&\int d\varphi_{\rm f}d\varphi_{\rm i} \,\frac{e^{-2U(\varphi_{\rm f})} }{\mathcal{N}}P(\varphi_{\rm i})\langle 0_{\rm out}| T\varphi(t_1)\ldots \varphi(t_n)|0_{\rm in}\rangle \nonumber \\
&=&  \langle 0_{\rm out}|T\varphi(t_1)\ldots \varphi(t_n)|0_{\rm in}\rangle \,.
\eea
Finally, we can define the generating functional $W[J]$ of equilibrium expectation values in the usual way
\beq\label{Generating 1}
Z[J]=\exp\left\{-W[J]\right\} = \int d\varphi_{\rm f} \,\frac{e^{-2U(\varphi_{\rm f}) + J_{\rm f} \varphi_{\rm f}}}{\mathcal{N}}  \langle 0_{\rm out}|0_{\rm in}\rangle^J
\eeq 
and obtain unequal time connected correlators via 
\bea
\frac{\delta^n W[J]}{\delta J(t_1)\ldots \delta J(t_n) }\Big|_{J=0}  &=& \frac{\langle 0_{\rm out}|T\varphi(t_1)\ldots\varphi(t_n)|0_{\rm in}\rangle}{\langle 0_{\rm out}|0_{\rm in}\rangle} \nonumber\\
&=& \langle 0_{\rm out}|T\varphi(t_1)\ldots\varphi(t_n)|0_{\rm in}\rangle
\eea
or late equal-time expectation values  
\beq
\frac{\delta^n W[J]}{\delta J_{\rm f}^n}\Big|_{J=0}  = \int d\varphi_{\rm f} \, \varphi_{\rm f}^n \,\frac{e^{-2U(\varphi_{\rm f})}}{\mathcal{N}}
=\langle \varphi_{\rm f}^n\rangle 
\eeq

\subsection{Effective mass and average effective action}

\begin{figure}[t]
	\begin{center}
		{\includegraphics[scale=0.6]{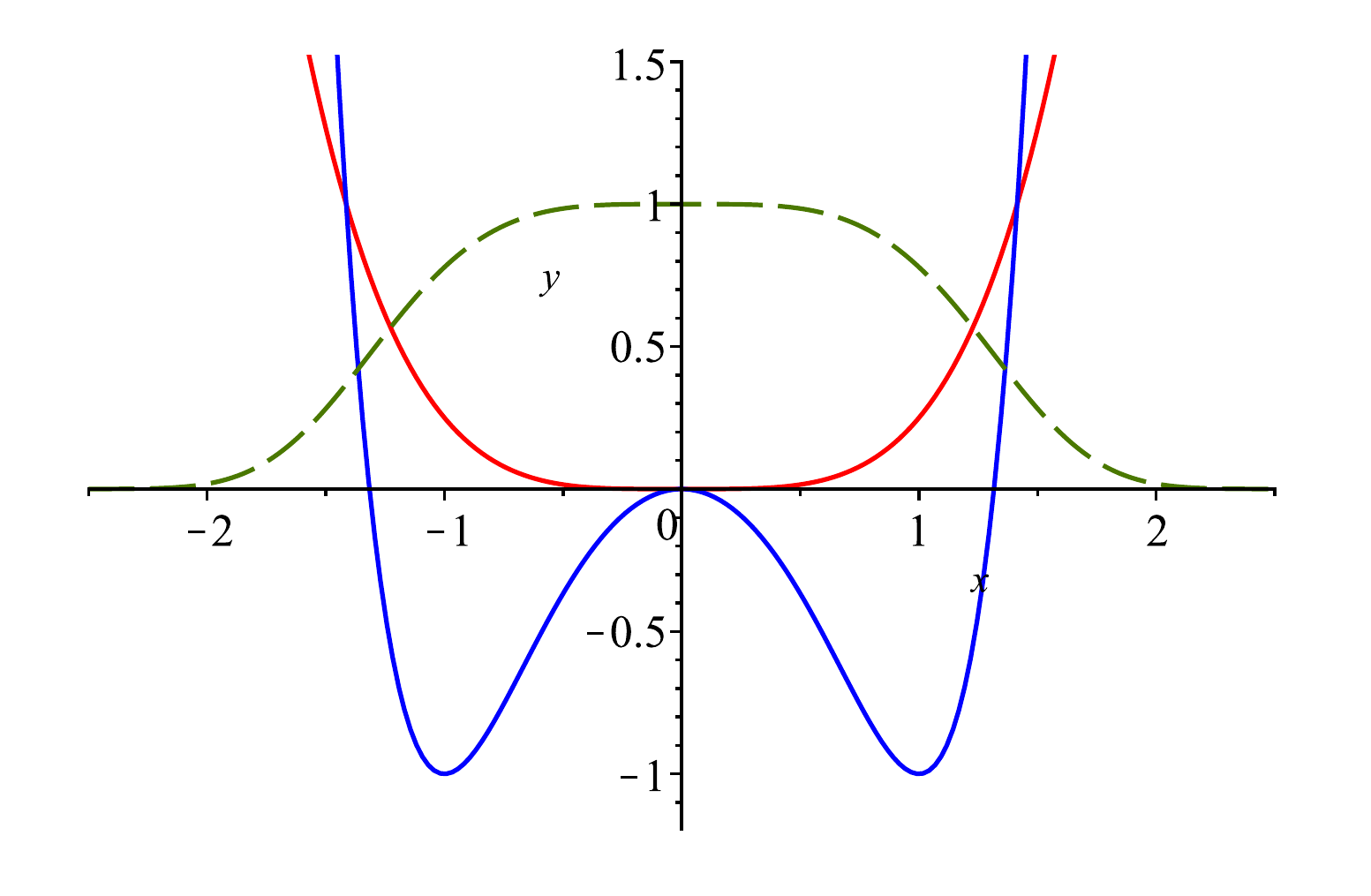}}
		\caption{A $U(\varphi)=\hat{\lambda} \varphi^4$ potential for the particle dynamics (red line) corresponds to the stochastic potential ${\cal W}(\varphi)= \frac{1}{2}\hat{\lambda}^{2}\varphi^6 - \frac{3}{2}\hat{\lambda}\varphi^2$ (blue line). The ground state, corresponding to the very late time limit and a zero eigenvalue is $\Psi_0(\varphi)=e^{-U(\varphi)}$, plotted here as the green dashed line. The equilibrium probability distribution is given by $P(\varphi)=e^{-2U(\varphi)}$. Time evolution towards this state is mostly determined by $\Lambda_1$, the lowest non-zero eigenvalue.  }
		\label{Ground state}
	\end{center} 
\end{figure}

As we mentioned above, the potential ${\cal W}$ determines the decay rate of unequal time correlators through the first eigenvalue $\Lambda_1$. In the language of the next section this will correspond to defining an effective mass for the field. For concreteness let us here focus on the massless self-interacting case   
\beq\label{potential}
V(\phi)=\frac{\lambda}{4}\phi^4\Rightarrow U(\varphi)= \frac{\hat{\lambda}}{4}\varphi^4\Rightarrow  {\cal W}(\varphi)= \frac{1}{2}\hat{\lambda}^{2}\varphi^6 - \frac{3}{2}\hat{\lambda}\varphi^2 \,,
\eeq
where $\hat{\lambda}=\lambda H^2/12\pi^2$. Rescaling $\varphi=y/\hat{\lambda}^{1/4}$ and $\mathcal{E}_n =\Lambda_n/\sqrt{\hat{\lambda}}$ we have from~(\ref{Schro-Euclidean}),
\beq
-\frac{\partial^2 \Psi_n}{\partial y^2}+\left(\frac{1}{2}y^6 - \frac{3}{2}y^2 \right)\Psi_n=\mathcal{E}_n\Psi_n \,.
\eeq
As mentioned above, the lowest eigenvalue is zero, $\mathcal{E}_0=0$, see also figure \ref{Ground state}. It is not difficult to numerically obtain the higher eigenvalues, the lowest of which is $\mathcal{E}_1=0.9677$ or \cite{Starobinsky:1994bd}
\beq\label{eigenvalue-1}
\Lambda_1=0.9677\sqrt{\hat{\lambda}}
\eeq    
which implies that the dominant long-time behaviour of correlators is 
\beq
\langle\phi(t+\Delta t)\phi(t)\rangle \propto \exp\left(-\Lambda_1t\right)\,.
\eeq 
Hence, according to stochastic theory, the decay rate of unequal time correlators is slightly slower than that implied by the dynamical mass $\hat{m}_{\rm dyn}^2=1.046 \sqrt{\hat{\lambda}}$ - see (\ref{GS mass}). Taking higher eigenvalues into account will provide a small correction to the decay law which is not exactly exponential but these corrections are beyond the scope of this paper.

Let us now examine the above result using the notion of the effective action $\Gamma$ and elaborate on the fact that this dynamical effective action handles time dependent correlators with the effective mass determining their decay rate. Consider the Legendre transform of the generating functional $W[J]$ 
\beq
\Gamma[\chi] = W[J] - \int dt J(t)\chi(t)\,,
\eeq
where $\chi$ is the average field defined through
\beq
\frac{\delta W[J]}{\delta J(t)} = \langle\varphi(t)\rangle \equiv \chi(t)
\eeq
and satisfying the average equation of motion 
\beq
\frac{\delta \Gamma[\chi]}{\delta \chi(t)} =J(t)\,.
\eeq  
The average unequal time correlation function is given by
\beq
\frac{\delta^2 W[J]}{\delta J(t)\delta J(t')}= \langle \varphi(t)\varphi(t')\rangle - \langle\varphi(t)\rangle \langle\varphi(t')\rangle \equiv G(t,t')-\chi(t)\chi(t') 
\eeq
and it is straightforward to show that 
\beq\label{eqn:green-1}
\int d\tau\frac{\delta^2\Gamma}{\delta\chi(t)\delta\chi(\tau)}\frac{\delta^2 W[J]}{\delta J(\tau)\delta J(t')} = \delta \left(t-t'\right) \,.
\eeq 

As we will see in the next section, $\Gamma$ is controlled by an RG flow that reaches an asymptotic IR fixed point at a finite value of the cutoff $k$. It therefore makes sense to use the asymptotic value of the effective potential to define an effective action which also describes dynamical quantities. We will be using the Ansatz, valid in our restricted version of the local potential approximation,
\beq
\Gamma[\chi]=\int dt \left[\frac{1}{2}\dot{\chi}^2  
+ {\cal V}\Big]\right]\,,
\eeq
where ${\cal V}$ is the effective potential reached by the RG flow at $k=0$ if one starts with the potential ${\cal W}$ at k=H. The dynamical equation for the average field therefore becomes     
\beq\label{average dynamics}
\frac{d^2\chi}{dt^2} - {\cal V}_{,\chi}=0
\eeq
and, accordngly, the two point function satisfies
\beq
\left[\frac{d^2}{dt^2} -  {\cal V}_{,\chi\chi}\right] G(t,t') = -\delta(t-t')\,.
\eeq
Assuming that the equilibrium point ${\cal V}_{,\chi}$ corresponds to the average field $\chi \rightarrow 0$ (we will see below that this is the case), comparing with ${\cal W}$ and writing ${\cal V}_{,\chi\chi}(\chi=0) = \hat{m}^4_{\rm IR}$ where $\hat{m}^2_{\rm IR}=m^2_{\rm IR}/3H$, we see that at equilibrium
\beq
G(t,t')\sim \Theta(t-t') \exp \left[-\hat{m}_{\rm IR}^2\left(t-t'\right)\right]\,.
\eeq  

These results also allow for the correlator (\ref{edge-correlator}) to be estimated. In general 
\bea
\langle \varphi_{\rm t},t|\varphi_{\rm t'},t'\rangle &=& \mathcal{N}^{-1} \exp\left\{-\Gamma\left[\chi_{\rm cl}\right]\right\}
\eea
where $\chi_{\rm cl}$ satisfies (\ref{average dynamics}) with boundary conditions $\chi(t)=\varphi_t$ and
 $\chi(t')=\varphi_{t'}$. It is then easily seen that 
\beq\label{on-shell action}
\Gamma[\chi_{\rm cl}]=\frac{1}{2}\left(\varphi_{\rm f}\dot{\chi}(t_{\rm f})
  - \varphi_{\rm i}\dot{\chi}(t_{\rm i})\right) - \frac{1}{2}\int dt 
\left[\chi_{\rm cl}({\cal W})'(\chi_{\rm cl}) - 2{\cal W}(\chi_{\rm cl}) \right]
\,.
\eeq 
As will be seen from the RG flow in the next section, the effective potential ends up being dominated by the fluctuation generated quadratic term, see e.g.~figure (\ref{couplings}). We can therefore reasonably approximate~(\ref{average dynamics}) by
\beq
\frac{d^2}{dt^2}\chi_{\rm cl} - \hat{m}^4_{\rm IR} \, \chi_{\rm cl}\simeq 0\,,
\eeq
again with boundary conditions $\chi(t)=\varphi_t$ and $\chi(t')=\varphi_{t'}$. The propagator can now be computed explicitly. Indeed
\beq
\chi_{\rm cl} (\tau) \simeq \frac{\varphi_{\rm t'}\sinh \left[\hat{m}_{\rm IR}^2\left(t-\tau\right)\right]+\varphi_{\rm t}\sinh \left[\hat{m}_{\rm IR}^2\left(\tau-t'\right)\right]}{\sinh \left[\hat{m}_{\rm IR}^2\left(t-t'\right)\right]}\,,  
\eeq  
the last term of (\ref{on-shell action}) vanishes and we obtain
\beq
\Gamma[\chi_{\rm cl}] \simeq \frac{\hat{m}_{\rm IR}^2}{2}\frac{\cosh \left[\hat{m}_{\rm IR}^2\left(t-t'\right)\right]\left(\varphi_{t}^2+\varphi_{t'}^2\right) - 2\varphi_{t}\varphi_{t'} }{\sinh \left[\hat{m}_{\rm IR}^2\left(t-t'\right)\right]}\,,
\eeq
from which we can compute
\bea
\langle \varphi_{\rm t},t|\varphi_{\rm t'},t'\rangle &\simeq& \mathcal{N}^{-1} e^{-\frac{m_{\rm IR}^2}{2}\left(\varphi_{t}^2+\varphi_{t'}^2\right)} \left(1 + 2\hat{m}_{\rm IR}^2\,\varphi_{t}\varphi_{t'}\,e^{-\hat{m}_{\rm IR}^2\left(t-t'\right)} \right)\\
\mathcal{N}&\simeq& \int\limits_{-\infty}^{+\infty} e^{-\hat{m}_{\rm IR}^2\varphi^2} d\varphi = \frac{\sqrt{\pi}}{\hat{m}_{\rm IR}}\,.
\eea      
We are therefore led to 
\beq
\langle\varphi_{t}\varphi_{t'}\rangle \simeq \langle \varphi_{t'}^2\rangle e^{-\hat{m}_{\rm IR}^2(t-t')}
\,,
\eeq
which shows that, for small average field values, 
the decay rate of the late time field correlator is governed by $\hat m_{\rm IR}^2$.

\section{Functional Renormalization Group flow}

We will now examine the long time behaviour of the stochastic system from the viewpoint of the Renormalization Group. Our discussion above has brought together the fact that the long wavelength effective theory for a spectator scalar in de Sitter is {\it Supersymmetric Euclidean Quantum Mechanics}, where the scalar potential $U(\varphi)$ plays the role of the superpotential. It is also possible to describe the system in terms of the stochastic potential $W$ which has a special form $\ref{W}$, related to the supersymmetry of the stochastic action. Therefore, in order to formulate the RG two possibilities exist within the Local potential approximation: either to explicitly respect the supersymmetry with the regulator or break it. We examine both below.

\subsection{Supersymmetric flow equation}
We first seek a supersymmetric formulation of the Renormalization Group. This has been achieved in \cite{Synatschke:2008pv} whose treatment we follow closely and where we refer the reader for details regarding the derivations below.\footnote{In this respect, see \cite{Canet:2006xu,Canet:2011wf} which also examine the relation between the supersymmetry of stochastic systems and the RG. } The starting point is the superspace incarnation of (\ref{Action with ghosts})
\beq
\mathcal{S} = \int dtd\theta d\bar{\theta}\left[\frac{1}{2}\Phi K \Phi +i U(\Phi)\right]
\eeq
where $\Phi$ is the superfield 
\beq
\Phi = \varphi +\bar{\theta}\psi + \bar{\psi}\theta  +  \bar{\theta}\theta F
\eeq 
and $K=\frac{1}{2}\left(D\bar{D} - \bar{D}D\right)$ with
\beq
D=i\partial_{\bar{\theta}}-\theta\partial_t\,,\quad \bar{D}=i\partial_{\theta}-\bar{\theta}\partial_t\,.
\eeq
The flow equation is obtained by adding an IR regulating term $\Delta \mathcal{S}_k$ to the action and computing the Legendre transform $\Gamma_k$, the scale dependent average effective action, of the regulated generating functional - see eg \cite{Berges:2000ew} for a comprehensive review of the formalism. The authors of \cite{Synatschke:2008pv} choose
\beq\label{DeltaS}
\Delta \mathcal{S}_k = \frac{1}{2} \int dtd\theta d\bar{\theta} \,\Phi R_k \Phi
\eeq
where $R_k=R_1\left(-\partial_t^2,k\right)+R_2\left(-\partial_t^2,k\right)K$ with $R_1$ and $R_2$ functions of the cutoff frequency $k$ and $-\partial_t^2$, or $\omega^2$ in frequency space. The functional RG equation then reads
\beq
\partial_k \Gamma_k = \frac{1}{2} {\rm STr}\left\{\left[\Gamma^{(2)}_k+R_k\right]^{-1}\partial_k R_k\right\}\,.
\eeq  
$\Gamma^{(2)}$ is the second functional derivative wrt the dynamical fields and ${\rm STr}$ denotes a supertrace. For our purposes it suffices to ignore $R_2$ and choose $R_1(k)$ a function of the cutoff $k$ only.  The regulating action term (\ref{DeltaS}) then simply involves
\beq
U(\varphi) \rightarrow  U(\varphi) + \frac{1}{2}\varphi R_1\varphi\,,
\eeq   
and the effective action is taken to be of the form 
\beq\label{Average Action with ghosts}
\Gamma_k=\int dt \, \left[\frac{1}{2}\dot{\varphi}^2- \bar{\psi}\left(\partial_t+U^k_{\varphi\varphi}\right)\psi +iFU^k_{\varphi}+\frac{1}{2}F^2\right]\,,
\eeq
where we have not labeled the classical fields differently to avoid proliferation of symbols. As discussed in \cite{Synatschke:2008pv}, after choosing a simple regulator $R_1(-\partial_t^2, k)=k$ the flow equation becomes  
\beq\label{RG flow}
\partial_kU_k = \frac{1}{4}\,\frac{1}{k+\partial^2_\varphi U_k}
\,. 
\eeq
We note that the above RG equation is identical to the flow equation derived in \cite{Guilleux:2015pma} - see eq. (23) in that work - if one sets $k = \frac{\kappa^2}{3H}$ and revert back to $\phi$ and $V$. It is remarkable that a smoothing in time of a spatially local version of the IR dynamics, the stochastic theory studied here and its supersymmetric incarnation, results in a flow equation for the effective potential which is identical to the one obtained from a smoothing of spatial scales in the context of QFT. We hope to provide better understanding of this correspondence in future work.

\subsection{Minimum non-pertubative mass and symmetry restoration in de Sitter}

Guillleux and Serreau \cite{Guilleux:2015pma} observe that the asymptotic value of the flow equation~(\ref{RG flow}) when $\lim\limits_{k\rightarrow 0} m_k^2$ for a massless bare theory ($m_H=0$) can be obtained via a static, effective action type relation
\beq\partial_{\varphi\varphi}U^{k\rightarrow 0}(\varphi=0)\langle \varphi^2\rangle =1\,,
\eeq
where brackets indicate an average over the equilibrium distribution (\ref{equilibrium distn}).  This leads to the definition of a dynamical mass 
\beq\label{GS mass}
m_{\rm dyn}^2 = \frac{\sqrt{2}}{4}\frac{\Gamma\left(\frac{1}{4}\right)}{\Gamma\left(\frac{3}{4}\right)}\frac{\sqrt{3}}{2\pi}\,\sqrt{\lambda_H}H^2 \simeq  1.046\, \frac{\sqrt{3}}{2\pi}\,\sqrt{\lambda_H}H^2\,.
\eeq 
More detailed numerical solutions have also been provided in \cite{Guilleux:2015pma} and \cite{Synatschke:2008pv}.

Here we pursue a simple polynomial approximation approximation to demonstrate from the RG perspective the emergence of an effective mass dependent on the coupling as $m^2\sim \sqrt{\lambda}H$. For this purpose we employ         
\beq\label{potential ansatz}
U_k=\hat{e}_k+\frac{\hat{m}_k^2}{2}\varphi^2 +\frac{\hat{\lambda}_k}{4}\varphi^4
\,,
\eeq 
where, $\hat{m}^2=m^2/3H$ and $\hat{\lambda}=\lambda H^2/12\pi^2$. Higher order couplings are generated by the flow  and can be kept, see eg \cite{Gonzalez:2016jrn}.

\begin{figure}[t]
	\begin{minipage}{7cm}
		{\includegraphics[scale=0.5]{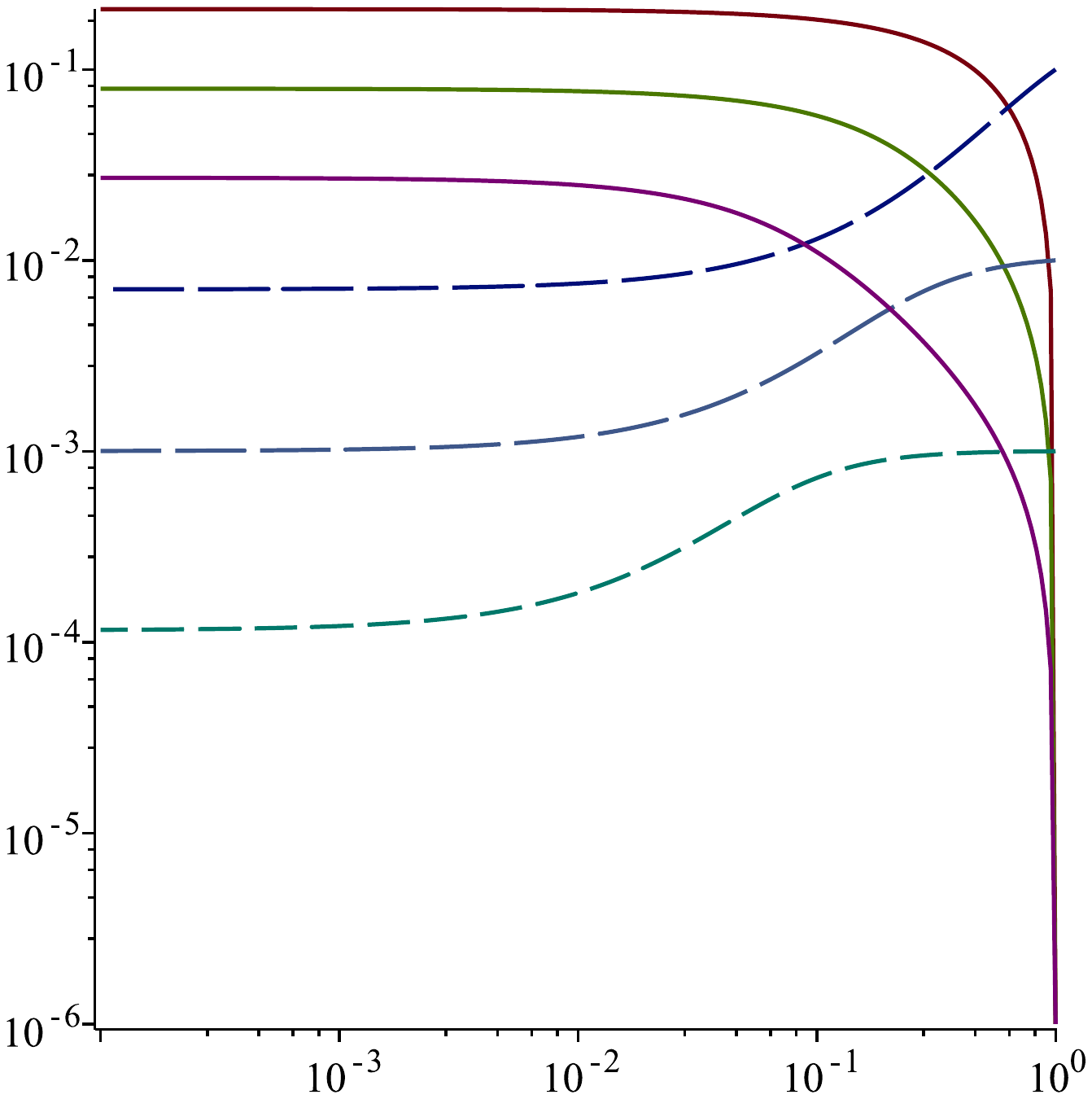}}
	\end{minipage}\hfill
	\begin{minipage}{7cm}
		\vspace{0.3cm}
		
		{\includegraphics[scale=0.5]{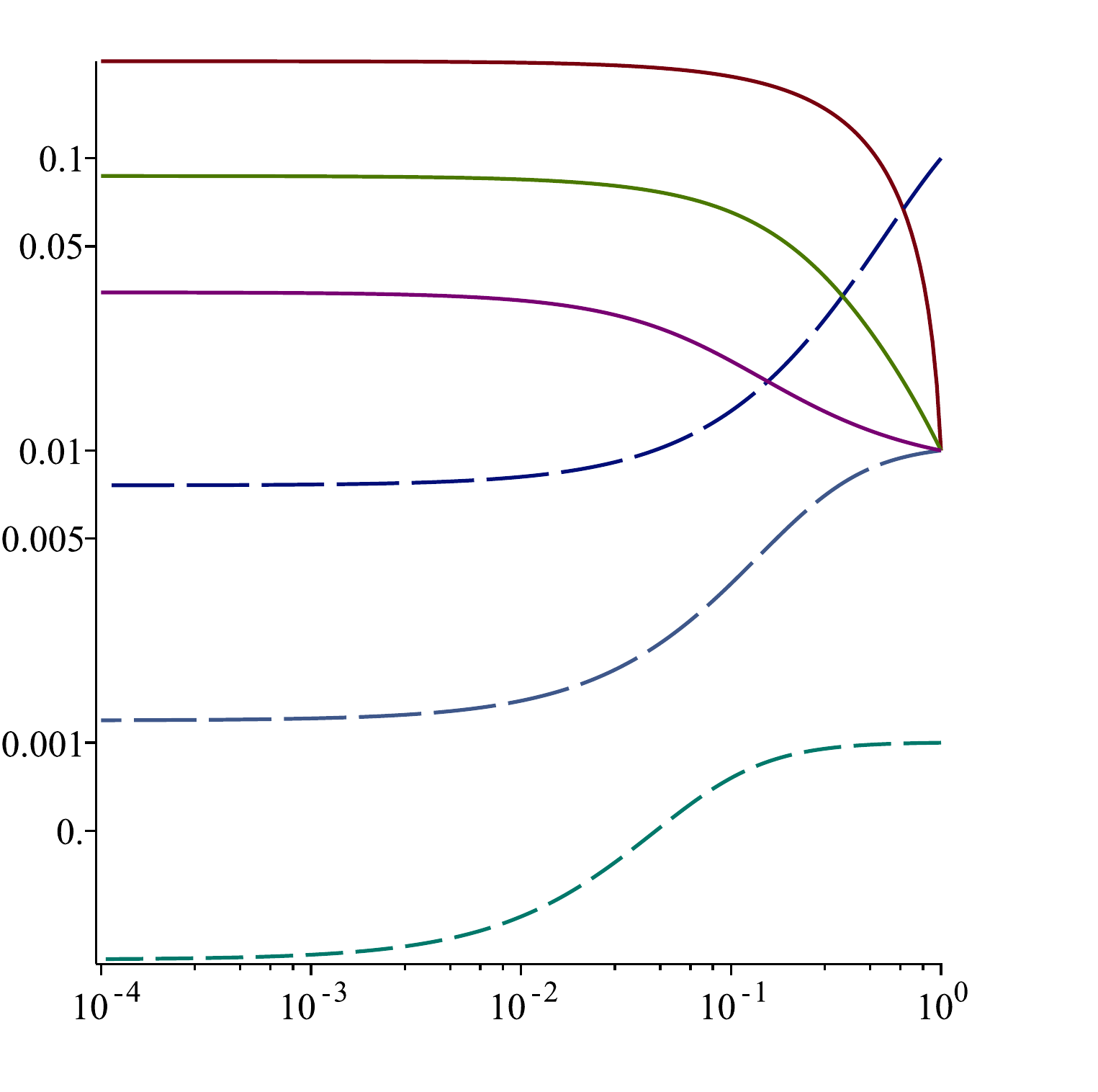}}
	\end{minipage}
	\caption{The flow of $\tilde{m}^2$ (solid lines) and $\tilde{\lambda}$ (dashed lines), shown in the vertical axis, with $k$ (horizontal axis) for an initial value $\tilde{m}^2_H=10^{-6}$ (left panel) and $\tilde{m}^2_H=10^{-2}$ (right panel). Three different initial couplings are shown $\tilde{\lambda}_H= 0.1,\, 0.01,\,0.001$ (top to bottom lines). The final values do not {\red show a} significant dependence on the initial value $\tilde{m}^2_H$.
	}
	\label{couplings} 
\end{figure}

Expanding $U_k$ on the rhs of (\ref{RG flow}) around $\varphi=0$ we obtain the flow equations 
\bea
\frac{d}{dk}\hat{e}_k&=&\frac{1}{4}\frac{1}{k+\hat{m}_k^2}\,,\\
\frac{d}{dk}\hat{m}^2_k&=&-\frac{3}{2}\frac{\hat{\lambda}_k}{\left(k+\hat{m}_k^2\right)^2}\,,\\
\frac{d}{dk}\hat{\lambda}_k&=&9\frac{\hat{\lambda}_k^2}{\left(k+\hat{m}_k^2\right)^3}
\,,
\eea
or, going to dimensionless variables $\hat{e}=\tilde{e}$, $\hat{m}^2=k\tilde{m}^2$ and $\hat{\lambda}=k^2\tilde{\lambda}$,
\bea\label{RG flow eqs}
k\frac{d}{dk}\tilde{e}_k&=&\frac{1}{4}\frac{1}{1+\tilde{m}_k^2}\,,\\\label{RG flow eqs2}
k\frac{d}{dk}\tilde{m}^2_k&=&-\tilde{m}_k^2-\frac{3}{2}\frac{\tilde{\lambda}_k}{\left(1+\tilde{m}_k^2\right)^2}\,,\\\label{RG flow eqs3}
k\frac{d}{dk}\tilde{\lambda}_k&=&-2\tilde{\lambda}_k+9\frac{\tilde{\lambda}_k^2}{\left(1+\tilde{m}_k^2\right)^3}
\,.
\eea
Numerical solutions for $\tilde{m}_k$ and $\lambda_{k}$ are shown in figure (\ref{couplings}).

\begin{figure}[t]
	\begin{minipage}{7cm}
		{\includegraphics[scale=0.5]{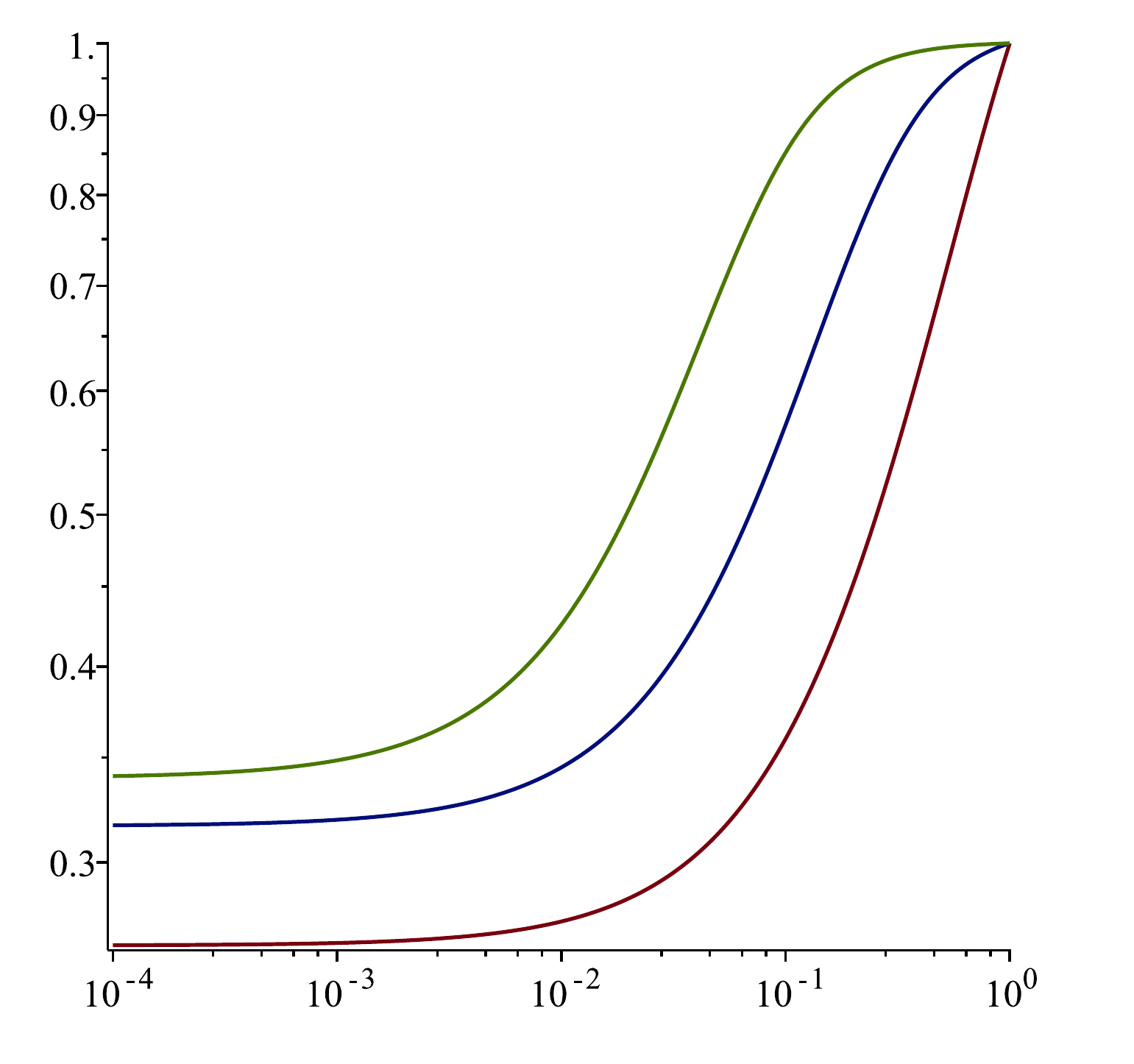}}
	\end{minipage}\hfill
	\begin{minipage}{7cm}
		{\includegraphics[scale=0.5]{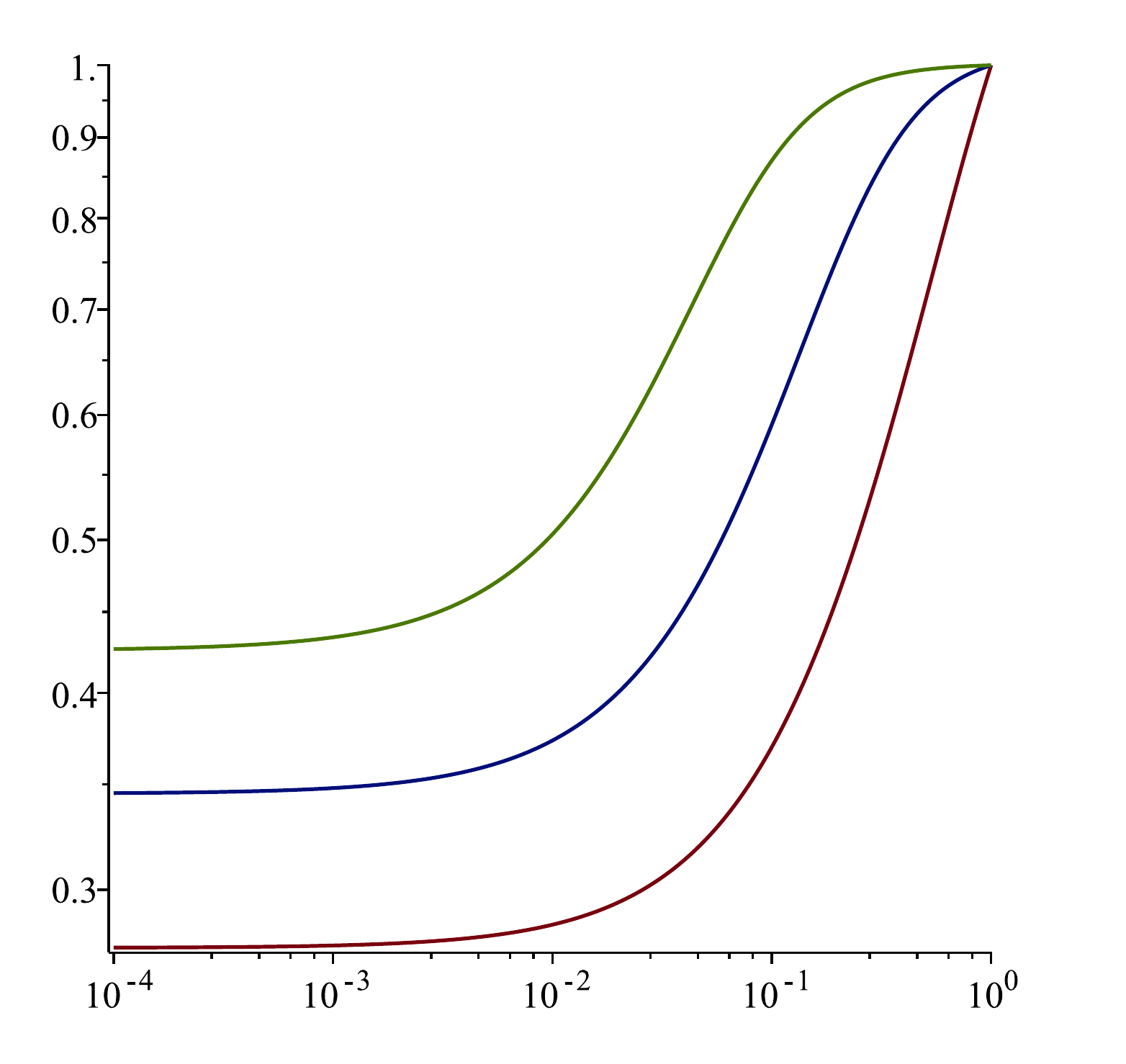}}
	\end{minipage}
	\caption{The ratio $\frac{\sqrt{\tilde{\lambda}_{k}}}{\sqrt{\tilde{\lambda}_{H}}}$ (vertical axis) for $\tilde{m}^2_H=10^{-6}$ (left panel) and $\tilde{m}^2_H=0.01$ (right panel) with $k$ (horizontal axis). The different lines, from top to bottom, correspond to $\tilde{\lambda}= 0.1,\, 0.01$ and $0.001$ respectively. We see that in the regime of interest the coupling weakens and this ratio drops below $\sim 0.45$. } 
	\label{coupling ratio} 
\end{figure}

\begin{figure}[t]\label{Flow}
	\begin{minipage}{7cm}
		{\includegraphics[scale=0.45]{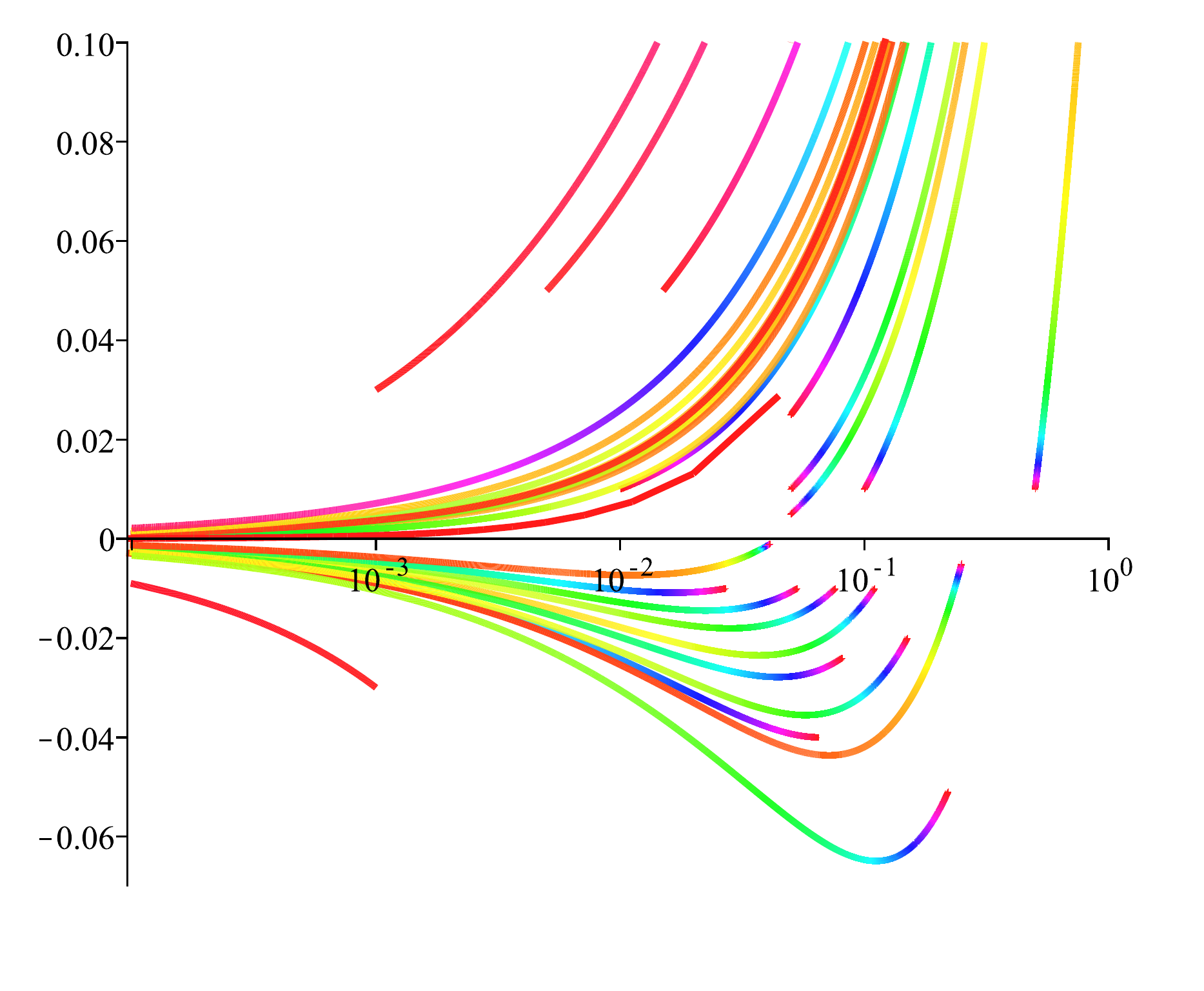}}
	\end{minipage}\hfill
	\begin{minipage}{7cm}
		{\includegraphics[scale=0.4]{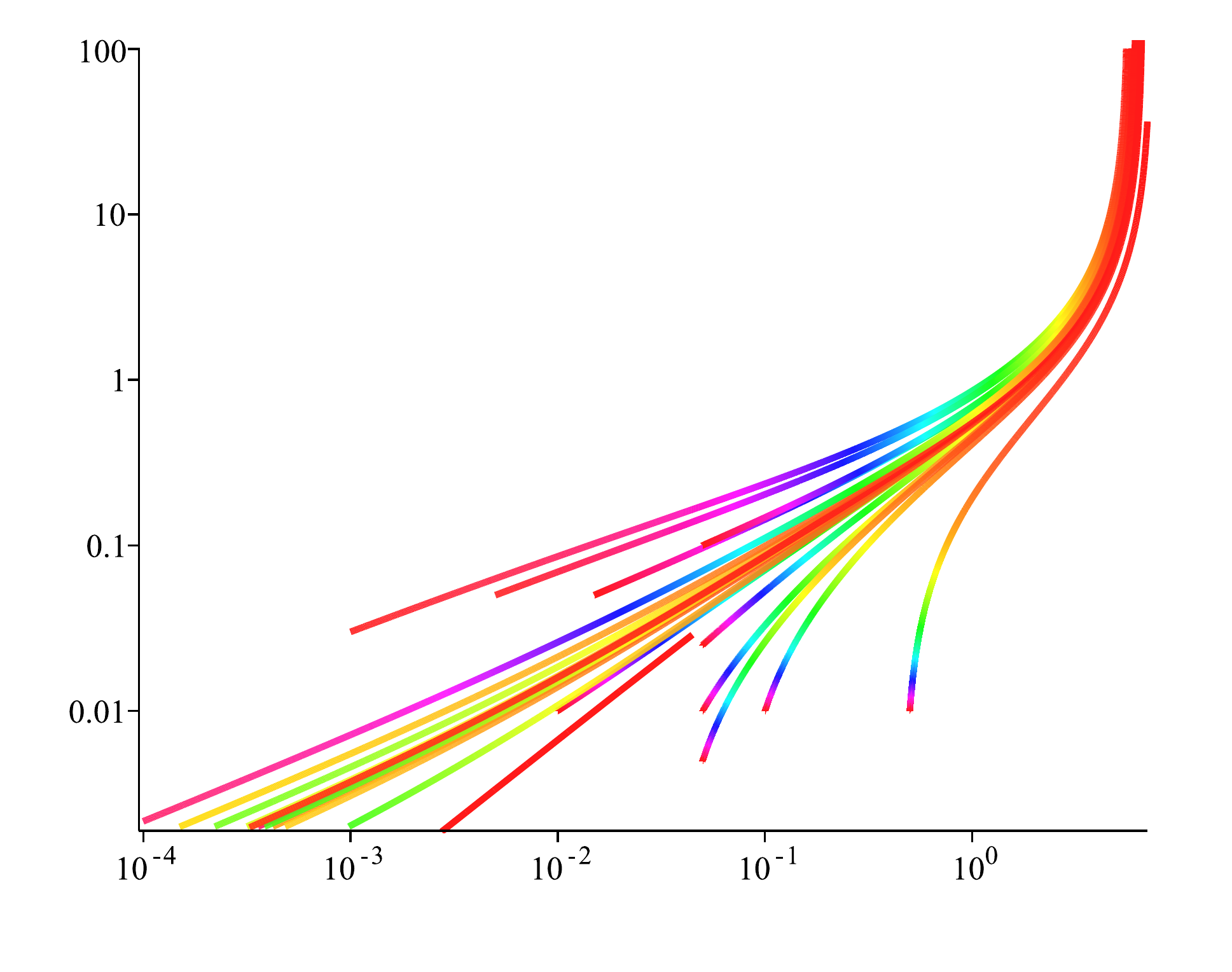}}
	\end{minipage}
\caption{The RG flow in the first quadrant of the $\tilde{m}_k^2-\tilde{\sigma}_k$ plane, with $\tilde m_k^2=m^2/(3Hk)$ (vertical axis) and 
$\tilde \sigma_k=4\pi^2 m^4/(3\lambda H^4)$ (horizontal axis). 
The colour scheme corresponds to $k$, the RG scale, with violet at the UV beginning of the flow at $k=H$ and red at $k=10^{-4} H$. 
The parameter $\tilde{\sigma}$ flows from a wide range of initial values with $|{\tilde{m}^2}|\ll 1$, $\tilde{\sigma}\ll 1$ to a constant value $\gamma \simeq 6$ which depends only weakly on initial conditions. An initially negative, symmetry breaking $\tilde{m}^2$ is transported through the origin, joining the flow in the first quadrant of the graph. The left panel is a closeup of the flow's initial portion while the right panel follows the flow in the first quadrant only. The attractor-like behaviour is evident. The trajectories shown to emerge from close to the origin in the right panel are those that crossed into the first quadrant from the symmetry breaking fourth quadrant of the $\tilde{m}_k^2-\tilde{\sigma}_k$ plane.             
		}
		\label{Flow}
\end{figure}

To illuminate the behaviour of the system further, let us define $\tilde{\sigma}=\tilde{m}_k^4/\tilde{\lambda}_k$. Convergence of the loop expansion (\ref{series}) is controlled by the parameter
\beq
\frac{3\lambda H^4}{4\pi^2 m^4}= \frac{\hat{\lambda}}{\hat{m}^4}=\tilde{\sigma}^{-1}\,.
\eeq    
For values $\tilde{\sigma}\lsim 1$ the higher loop expansion for the late time field variance appears invalid. The RG shows that this situation cannot in fact be realized in nature  --
there cannot exist 
such massless/light scalar fields in de Sitter: as we will see a $\tilde{\sigma}\lsim 1$ will be driven to values $\tilde{\sigma}\rightarrow {\rm const} \gtrsim 1$. Hence the field necessarily obtains a mass of the  $m^2\sim\sqrt{\lambda}H^2$, through self coupling to infrared fluctuations. This means that the deep IR physics of de Sitter is necessarily non-perturbative, and that fundamental excitations are not the free particles of the UV.

The RG equations (\ref{RG flow eqs2}) and (\ref{RG flow eqs3}) give
\bea
k\frac{d}{dk}\tilde{m}^2_k&=&-\tilde{m}^2_k\left(1+\frac{3}{2}\frac{\tilde{m}^2_k}{\left(1+\tilde{m}_k^2\right)^2}\frac{1}{\tilde{\sigma}_k}\right)\\
k\frac{d}{dk}\tilde{\sigma}_k&=&-3\frac{\tilde{m}^2_k\left(1+4\tilde{m}_k^2\right)}{\left(1+\tilde{m}_k^2\right)^3}
\,.
\eea 
Suppose that the "bare" theory at the UV cutoff $k=H$ has $\tilde{\sigma}\ll 1$. The flow equations are, approximately  

\beq
k\frac{d}{dk}\tilde{m}_k^2\simeq -\frac{3}{2}\frac{\tilde{m}_k^4}{\tilde{\sigma}_k}\,,\quad 
k\frac{d}{dk}\tilde{\sigma}_k\simeq -3\tilde{m}_k^2
\eeq 
from which we obtain
\beq\label{early flow}
\tilde{\sigma}_k^{1/2}=\tilde{\sigma}_H^{1/2}+\frac{3\alpha}{2}\ln\left(\frac{H}{k}\right)\,,\quad \tilde{m}_k^2=\alpha\tilde{\sigma}_k^{1/2}
\eeq
where $\alpha$ {\red is } a constant related to the bare value of $\lambda$ at the UV cutoff $k=H$ \beq
\alpha=\frac{\sqrt{\lambda_H}}{2\sqrt{3}\pi}
\eeq 
We see that both $\tilde{\sigma}_k$ and $\tilde{m}_k^2$ grow from their initial values at $k=H$. This trend continues and the flow is eventually described by 
\beq
k\frac{d}{dk}\tilde{m}_k^2\simeq - \tilde{m}_k^2 \,,\quad k\frac{d}{dk} \tilde{\sigma}_k\simeq-\frac{12}{\tilde{m}_k^2}
\eeq   
At late RG time 
\beq
\tilde{m}_k^2\simeq \beta k^{-1}\,,\quad \tilde{\sigma}_k=\gamma-\frac{12}{\beta}k^2
\eeq
where $\beta={\rm const}$ and ${\gamma=\rm const}$. A numerical investigation of the flow reveals that indeed $\hat{\sigma}$ approaches a constant value that is close to $\gamma \simeq 6$, for a wide range of initial conditions (provided the field remains light, {\it i.e.} 
 $m\ll H$),
see figure \ref{Flow}. Note however that, although the above 
behavior is strongly attractive, the exact value of $\gamma$ that a trajectory approaches depends somewhat on the initial values of $m^2$ and $\lambda$ at $k=H$, making $\gamma$ a ``quasi-universal'' parameter. Nevertheless, we can see that the field develops a mass parametrically dependent on $\sqrt{\lambda}$; as $k\rightarrow 0$ 
\beq
m^2_{k\rightarrow 0}=3\beta H\,,\quad \lambda_{k\rightarrow 0} =12\pi^2 \frac{\beta^2}{\gamma H^2}
\eeq
with the advertised relation 
\beq\label{IR mass}
m^2_{k\rightarrow 0}=\frac{\sqrt{3\gamma}}{2\pi}\sqrt{\lambda_{k\rightarrow 0}}\,H^2=\sqrt{\gamma}\frac{\sqrt{\lambda_{k\rightarrow 0}}}{\sqrt{\lambda_{H}}}\frac{\sqrt{3}}{2\pi}\,\sqrt{\lambda_H}H^2 \,. 
\eeq 
The simple approximation to the RG flow adopted in this paper gives a slightly varying range of results for a wide range of different bare theories. For the parameter $\gamma$ we have $\gamma\simeq 6-7$ depending weakly on the initial condition satisfying $m^2 < 1$ and $\sigma < 1$. For trajectories starting close to $m^2_H=0$, the values of  $\sqrt{\lambda_{k\rightarrow 0}/\lambda_H}$ cluster closely around $\sim 0.3$ while $\gamma \simeq 6.2$, depending on the exact value of the initial coupling $\lambda_{H}$. This gives
\beq
m_{\rm IR}^2 \simeq 1.29 \, \frac{\sqrt{3}}{2\pi}\,\sqrt{\lambda_H}H^2 \,.
\eeq
We see that the very simple quartic truncation to the flow equations used here does a decent job in reproducing the exact result (\ref{GS mass}) to about 20\%.

\begin{figure}[t]
	\begin{center}
		{\includegraphics[scale=0.6]{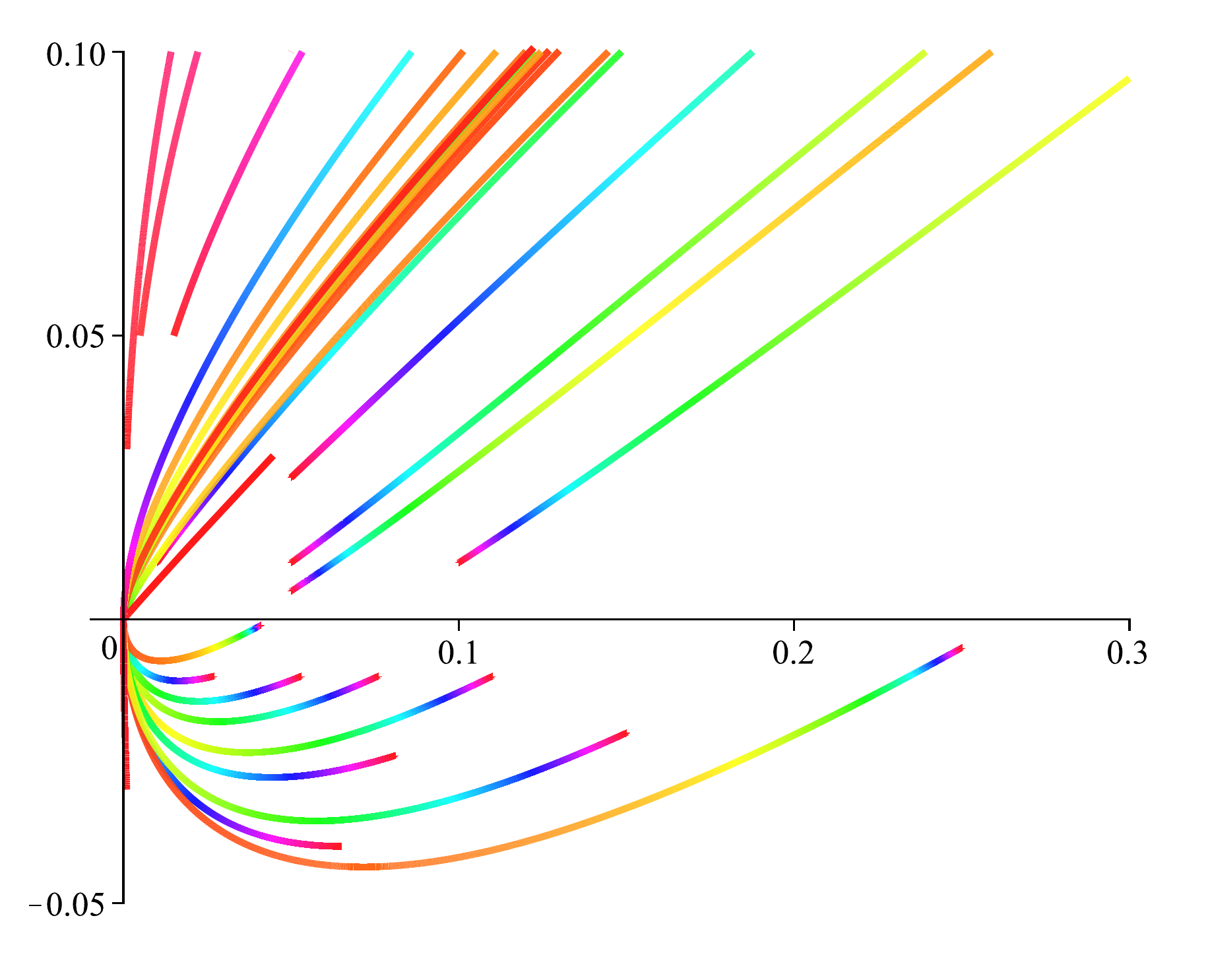}}
		\caption{Same as in the left panel of figure \ref{Flow} but with a linear $\tilde{\sigma}$ axis, showing trajectories crossing from the symmetry breaking into the positive mass quadrant.      
		}
		\label{OriginLinear}
	\end{center} 
\end{figure}
The above RG equations allow us to also probe the phenomenon of symmetry restoration in de~Sitter \cite{Serreau:2011fu, Lazzari:2013boa}. Indeed, starting at the $4^{th}$ quadrant of the $\tilde{m}_k^2-\tilde{\sigma}_k$ plane, corresponding to symmetry breaking $\tilde{m}^2<0$, the flow passes through zero into the first quadrant where it joins the flow described above. Note that at $(\tilde{m}^2,\tilde{\sigma})=(0,0)$ the ratio $\tilde{m}^4/\tilde{\sigma}$ remains finite, being simply the coupling constant. Therefore, all trajectories from the $4^{th}$ quadrant join trajectories in the $1^{st}$ quadrant through the origin with different values of $\alpha$ in (\ref{early flow}). These trajectories then converge to a constant value for $\tilde{\sigma}$, see figures \ref{Flow} and \ref{OriginLinear}.

\subsection{Non-supersymmetric flow equation}
Let us now consider the stochastic action in the form  (\ref{stochastic Gen}), after the ghost and auxiliary fields have been integrated out, and directly add a regulating term  
\beq
\Delta S_k= \frac{1}{2}\int\frac{d\omega}{2\pi} \varphi(-\omega)R_k(\omega)\varphi(\omega)\,.
\eeq
The Wetterich equation for the average effective action $\Gamma_k$ (here $\partial/\partial t = k\partial/\partial k$)
\beq
\frac{\partial}{\partial t}\Gamma_k[\varphi] = \frac{1}{2}{\rm Tr}\left[\partial_tR_k\left(\Gamma_k^{(2)}\left[\varphi\right]+R_k\right)^{-1}\right]\,,
\eeq
with the LPA ansatz 
\beq
\Gamma[\varphi]=\int dt \left[\frac{1}{2}\dot{\varphi}^2  + V_k(\varphi)\right]\,
\eeq 
and a regulator
\beq
R_k(\omega)=\left(k^2-\omega^2\right)\Theta\left(k^2-\omega^2\right)
\eeq
leads to the flow equation 
\beq
\frac{d}{d k} V_k(\varphi)= \frac{1}{\pi}\frac{k^2}{k^2+\partial^2_{\varphi}V_k(\varphi)}\,
\eeq
with the initial condition $V_H(\varphi)=W(\varphi)$.  
\begin{figure}[t]
	\begin{center}
		{\includegraphics[scale=0.6]{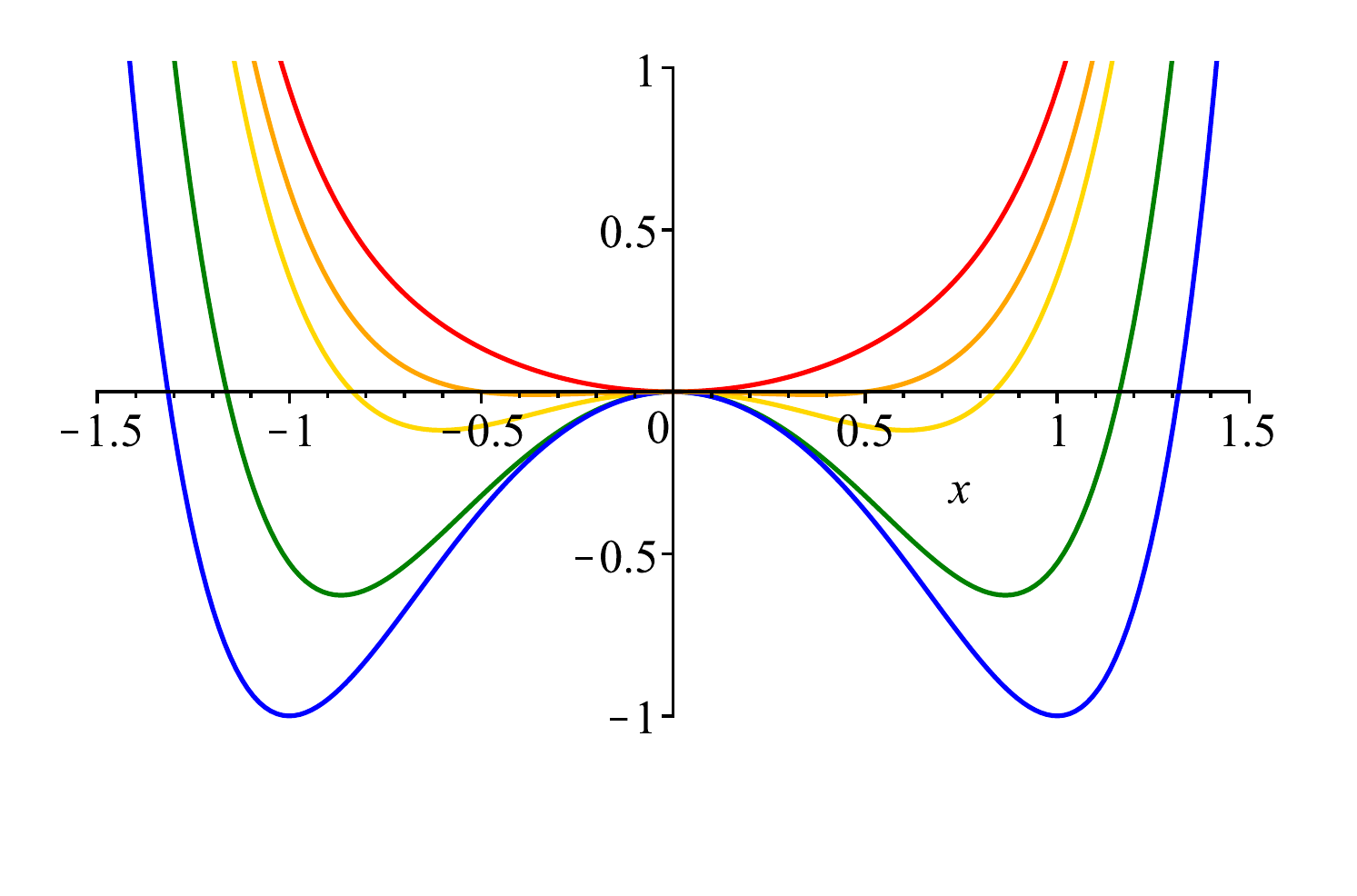}}
		\caption{The non-supersymmetric flow of the stochastic potential, initially defined at $k=H$ by $W(\varphi)$, shown here as the lower, blue line.}
		\label{non-Susy flow}
	\end{center} 
\end{figure}
Numerical solutions of this equation are shown in figure \ref{non-Susy flow}, again clearly demonstrating the development of an effective positive mass. Its numerical value is computed to be 
\beq
\hat{m}^2_{\rm IR}= 1.010 \sqrt{\hat{\lambda}_H}\Rightarrow m_{\rm IR}^2 =1.010 \frac{\sqrt{3}}{2\pi}\sqrt{\lambda_H}H^2\,,
\eeq 
a result very close to (\ref{eigenvalue-1}). The analysis described in \cite{Canet:2006xu, Canet:2011wf} implies the importance of supersymmetry in the stochastic system yet a reasonable approximation to the mass is obtained even by a flow that breaks it. It would be interesting to further examine whether this carries any significance in future work.

\section{Summary and Discussion}
The local value of a light scalar $\phi$ in de~Sitter, averaged over a patch of size $\Delta r \sim H^{-1}$,  obeys stochastic Langevin dynamics. We applied the functional renormalization group to this theory in order to understand \emph{temporal} fluctuations over timescales large compared to $H^{-1}$. We found that integrating them out generates a mass that damps temporal correlators, curing associated secular (infrared) divergences, and restores the symmetry in potentials that break it. Formally, the deep IR effective mass squared we have calculated here is (up to a volume factor) the zero momentum limit of the two point proper vertex, which is the inverse of the full two point function.

Remarkably, the flow equation obtained with a supersymmetric regulator is identical to that obtained in previous works utilizing the functional RG to the full QFT in de~Sitter and smoothing spatial fluctuations \cite{Serreau:2011fu, Serreau:2013eoa,Guilleux:2015pma, Guilleux:2016oqv, Kaya:2013bga}. The reason for this agreement can probably be traced to the underlying de Sitter invariance of the infrared correlators~\cite{Gautier:2013aoa,Garbrecht:2013coa}. 
Therefore, the results obtained previously for the flow with spatial smoothing can be transported to the temporal domain. A particular example is symmetry restoration. In the temporal domain it can simply be seen as the result of strong fluctuations transporting the field over the potential barrier such that, when smoothed over sufficiently long timescales, the barrier is not relevant for the dynamics any more. One should note that 
invoking de Sitter invariance to relate large temporal and spatial correlators is not as trivial as it may appear,
since de Sitter symmetry is not apparent in stochastic inflation and moreover different coarse graining procedures 
may violate it. Furthermore, not all interacting theories have a well defined de Sitter invariant 
deep infrared physics, notable examples are Yukawa theopy~\cite{Miao:2006pn} and possibly quantum gravity.

Our discussion has largely utilized the supersymmetry of the effective stochastic IR action of a light scalar in de Sitter. It would be interesting to examine if and to what extent the stochastic nature of the IR dynamics  is another example of an emergent IR supersymmetry such as that recently studied in \cite{Gies:2017tod}.                 

\section*{Acknowledgements}
GR would like to thank Ian Moss for useful conversations and Julien Serreau for sharing his knowledge on the applications of the Functional Renormalization Group to scalars in de Sitter and extensive discussions on the work presented in this paper. The authors would also like to thank the anonymous referee whose substantial input has allowed us to improve this paper significantly. GR is partially supported by STFC grant ST/P000371/1 - Particles, Fields and Spacetime.

\section{Appendix}
In this appendix we briefly recall for completeness the path integral formulation of systems described by a Langevin equation, known as the MSRJD path integral. The discussion follows \cite{Garbrecht:2014dca}; for more details see the e.g. the textbooks \cite{Kamenev, Calzetta:2008iqa}. Consider the stochastic Langevin equation
\beq
\dot{x}+V'(x)=\xi(t)
\eeq
where $V(x)$ is a potential function and a prime denotes $d/dx$.  The noise is Gaussian and white, such that
it is fully determined by its two point correlation function, which reads
\beq
\la \xi(t)\xi(t^\prime) \ra = A\delta(t-t^\prime)\,,
\label{eq:noise_correlation}
\eeq
where the expectation value $\la\dots\ra$ denotes the average over noise realizations. Expectation values of any functional $\mathcal{O}[x]$ w.r.t.~the different realizations of the stochastic force $\xi$ can be obtained from the following path integral

\bea
\langle \mathcal{O} [x] \rangle &=& %\frac{1}{\mathcal{Z}}
\int D\xi {\rm e}^{- \frac{1}{2} \int dt \, \frac{\xi^2}{A} }  \int Dx \, \mathcal{O}[x]\delta(x-x[\xi])\nonumber \\
&=&\int D\xi {\rm e}^{- \frac{1}{2} \int dt \, \frac{\xi^2}{A} }  \int Dx \, \mathcal{O}[x] \, \delta \left[\dot{x} + V'  - \xi\right] \, \mathcal{J}[x]\,,\label{eq:stoch-path}
\eea
\noindent where $\delta[\ldots]$ denotes the Dirac delta functional and $\mathcal{J}[\phi]$ is the Jacobian of the argument of the delta function with respect to the integration variable $x$: $\mathcal{J}[x]= \left|{\rm Det}\left[\frac{\delta}{\delta x}\left(\dot{x} +\partial_{x} V - \xi\right)\right]\right| = \left|{\rm Det}\left[\partial_t + V''\right]\right|$. The value of the determinant depends on the discretization of the $\partial_t$ operator in the definition of the path integral or, equivalently in the continuum notation, the value assigned to $\Theta(0)$, the step function at the origin. The integration over the noise $\xi$ reflects the assumption that the latter is Gaussian. Expressing the delta functional via its Fourier transform and doing the integral over $\xi$ we arrive at  
\beq\label{eq:partition 1}
\langle O[x]\rangle = \int Dx D\psi \,O[x]\,\mathcal{J}[x]\,{\rm e}^{- \int dt \left\{i \psi \left(\dot{x} + V'\right) + \frac{A}{2} \psi^2\right\}}\,.
\eeq
To directly obtain expectation values like the above, it is convenient to define the partition function 
\beq\label{eq:partition 2}
\mathcal{Z}\left[J_x,J_\psi\right] = \int Dx D\psi \,\mathcal{J}[x]\,{\rm e}^{- \int dt \left\{i \psi \left(\dot{x} + V'\right) + \frac{A}{2} \psi^2 + i J_xx +iJ_\psi\psi\right\}}\,.
\eeq
from which $n$-point functions can be computed in the usual way by taking derivatives with respect to $-{\rm i}J_\psi$ and $-{\rm i}J_\phi$. Note that, unlike the standard QFT partition function, $\mathcal{Z}[0,0]=1$, and expectation values can be obtained directly from $\mathcal{Z}$ without receiving multiplicative contributions from vacuum bubbles.

In order to prepare for a perturbative expansion, we replace
$\dot x\psi \to \frac12( \dot{x}\psi - x\dot{\psi})$ under the
integral, and we
decompose the potential, including the contribution from exponentiating the Jacobian determinant, as $V = \frac{1}{2}m x^2 + V_{\mathrm{int}}(x)$. This
yields
\bea
\label{def:go1}
\mathcal{Z}\left[J_x,J_\psi\right] &=& \int D x D \psi \,\, {\rm e}^{- i\int dt \left\{ \frac{1}{2} \bmatr x, & \psi \ematr \bmatr 0 & (-\partial_t + m) \\  (\partial_t + m) & -{i}A \ematr \bmatr x \\ \psi \ematr + \bmatr J_x, & J_\psi\ematr \bmatr x\\ \psi\ematr  + V^{\prime}_{\mathrm{int}}\psi \right\}}
\\
\notag
&\equiv& \int Dx D\psi \,\, {\rm e}^{- i\int dt \left\{ \frac{1}{2} \bmatr x, & \psi \ematr \mathbb{G}_0^{-1} \bmatr x \\ \psi \ematr + \bmatr J_x, & J_\psi\ematr \bmatr x\\ \psi\ematr +  V^\prime_{\mathrm{int}}\psi \right\}},
\eea
where the last equality defines $\mathbb{G}_0^{-1}$. The latter is the functional and matrix inverse of the free propagator $\mathbb{G}_0$, that is given by
\beq
\mathbb{G}_0(t,t^\prime) = \bmatr \la x(t)x(t^\prime) \ra & \la x(t)\psi(t^\prime) \ra \\ \la\psi(t)x(t^\prime) \ra & \la \psi(t)\psi(t^\prime) \ra  \ematr \equiv \bmatr F(t,t^\prime) & -\mathrm{i}G^R(t,t^\prime) \\ -\mathrm{i}G^A(t,t^\prime) & 0  \ematr\,,
\label{def:go}
\eeq
where the above equality defines the free propagators $G^{R,A}$ and $F$. It should be emphasized that the null entry in $\mathbb{G}_0$ is a direct consequence of the definition of $\mathbb{G}_0^{-1}$.
It occurs due to the fact that $\psi$ is an auxiliary field and therefore not dynamical. Using Eqs.~(\ref{def:go1},\ref{def:go}) and the relation $\mathbb{G}_0 \star \mathbb{G}_0^{-1} (t,t^\prime) = \mathbb{I}_{2\mathrm{x}2}\delta(t-t^\prime)$,
we observe that $G^{R,A}$ are the retarded and advanced propagators for the operator $(\partial_t + m)$, while the statistical correlator is the two-point function of the original field $F(t,t^\prime) = A\int d\tau G^R(t,\tau) G^R(\tau,t^\prime) + F_0(t,t') $,
%{\red $F(t,t^\prime) = \la \frac12\{x(t),x(t^\prime)\} \ra $} {\blue To relate to the anti-commutator one would need to specify the relation with Schwinger - Keldysh.}
where $F_0(t,t')$ is an arbitrarily chosen solution to the equation of motion $\left(\partial_t+m\right)F_0(t,t')=0$. These Green functions can be easily found, and they read
\begin{subequations}
	\label{free:propagators}
	\bea
	G^R(t,t^\prime) &=& G^A(t^\prime,t) = {\rm e}^{-m(t-t^\prime)}\Theta(t-t^\prime), \label{def:gret_stoc} \\
	F(t,t^\prime) &=& \frac{A}{2m}\left(e^{-m|t-t^\prime|} -  e^{-m(t+t^\prime)} \label{def:F_stoc} \right) + F_0(t,t')\,,\nonumber \\
	&=& \frac{A}{2m} e^{-m|t-t^\prime|}\,.
	\eea
\end{subequations}
Notice that the otherwise arbitrary $F_0(t,t')$ was chosen here to make $F(t,t') = F(t-t')$. This fixes $F_0(t,t')$ uniquely. The above elements along with an appropriate interaction vertex can be used for a Feynman diagram expansion of correlation functions - see \cite{Garbrecht:2013coa, Garbrecht:2014dca, Petri:2008ig}.

\end{document}